\documentclass[aps,pre,notitlepage]{revtex4-1}
\usepackage{color,bm,epsfig,graphics,amssymb,amsmath,subeqnarray,setspace,graphicx,amsthm,epstopdf,subfigure,color}
\usepackage{float}
\def\e{\varepsilon}
\def\b{\mathbf}

\def\U{\b{U}}
\def\O{\bm{\Omega}}
\def\Uo{\b{U}_0 } 
\def\Oo{\bm{\Omega}_0 }

\def\o{\hat{\Omega}}

\def\Atens{\mathsf{A}}
\def\Btens{\mathsf{B}}
\def\Dtens{\mathsf{C}}

\def\D{\mathsf{D}}
\def\I{\mathsf{I}}

\def\r{\b{r}}
\def\rp{\dot{\b{r}}}
\def\rpp{\ddot{\b{r}}}
\def\rppp{\dddot{\b{r}}}

\newcommand\tcb{\textcolor{black}}

\begin{document}

\title{Helical trajectories of swimming cells with a flexible flagellar hook}

\author{Zonghao Zou}
\author{Wilson Lough}
\author{Saverio Spagnolie}
\email[]{spagnolie@math.wisc.edu}
\affiliation{Department of Mathematics, University of Wisconsin--Madison, 480 Lincoln Drive, Madison, Wisconsin 53706, USA}

\date{\today}

\begin{abstract}
The flexibility of the bacterial flagellar hook is believed to have substantial consequences for microorganism locomotion. Using a simplified model of a rigid flagellum and a flexible hook, we show that the paths of axisymmetric cell bodies driven by a single flagellum \tcb{in Stokes flow} are generically helical. Phase-averaged resistance and mobility tensors are produced to describe the flagellar hydrodynamics, and a helical rod model which retains a coupling between translation and rotation is identified as a distinguished asymptotic limit. A supercritical Hopf bifurcation in the flagellar orientation beyond a critical ratio of flagellar motor torque to hook bending stiffness, which is set by the spontaneous curvature of the flexible hook, the shape of the cell body, and the flagellum geometry, can have a dramatic effect on the cell's trajectory through the fluid. Although the equilibrium hook angle can result in a wide variance in the trajectory's helical pitch, we find a very consistent prediction for the trajectory's helical amplitude using parameters relevant to swimming {\it P. aeruginosa} cells.
\end{abstract}

\maketitle

\section{Introduction} 

One of the primary means of prokaryotic microorganism propulsion in viscous fluids is through the rotation of one or many helical flagella, and each flagellum is connected to a rotary motor embedded in the cell membrane by a flexible elastic hook which acts as a joint \cite{Lighthill76,bw77,macnab76,s04}. The bending stiffness of flagellar hooks varies widely, from $\sim10^{-4}$ pN $\mu$m$^2$ to $\sim10^{-1}$pN $\mu$m$^2$ \cite{bbb89,bbb91,sng04,fm04,pblbba09,sgs13}, with smaller values in peritrichous (multi-flagellated) organisms which require the increased flexibility so that the flagella may form flagellar bundles \cite{bsswdloab12}; the hook length itself may be optimized in nature in service to helical bundle stability \cite{sporing2018hook}. The stiffer hook appearing in monotrichous organisms (which propel using a single polar flagellum) appears necessary to stabilize straight swimming, but hook compliance is also needed during the ``flick'' phase of so-called ``run-reverse-flick'' trajectories in {\it V. alginolyticus} during chemotaxis \cite{kinsm05,xacw11,sgs13,pkl19}. Although it has been observed in generic settings that a helical flagellum does not substantially {\it deform} under rotation \cite{dtrb07}, a more recent investigation by Jabbarzadeh \& Fu indicates that both hook as well as flagellum deformability is needed to account for the large hook angles seen in such flicks \cite{jf18,jf20b}, consistent with experimental observations \cite{xacw11}. Even without the added complexity of a cell body, the chirality of a helical filament results in coupling of translation and rotation which can lead to surprisingly rich dynamics under gravity \cite{pdbm18}, under magnetic actuation \cite{cy19,smmf20}, near surfaces \cite{dl19,Ishimoto19}, in a background flow \cite{ishimoto20,zcb21} or even double-helical trajectories for double-helical ``superhelices'' like insect spermatozoa \cite{jm07,psl12}. For very soft filaments, other instabilities and dynamics abound \cite{pmclr15,jkdgr15,cllcfrsl20}.

The end result of such flagellar activity is the body trajectory, itself an object of intense scrutiny. Observations of the helical paths of swimming microorganisms date back as far as the eloquent descriptions by Jennings in 1901 \cite{Jennings1901}. The helical trajectories of microorganisms have been explored in a very general setting and shown to be demanded by differential geometry under the assumptions of a fixed velocity in the body frame by Crenshaw \cite{c931}, who used the criterion that a path with fixed curvature and fixed torsion is sufficient to identify a perfect helical trajectory. Helical swimming is a natural consequence when the driving flagellum is tilted at an angle relative to the surface at its connection point to the cell body. In addition to providing a thrust force on the cell, a tilted flagellum, rotated at its base by the rotary motor, will precess in a circular motion relative to the body surface. Though apparently detrimental to motility, this bending angle and precession in fact leads to enhanced mobility in {\it C. crescentus} \cite{lgmtpb14}. The ``wobbling'' and ``wiggling'' of cell bodies has been investigated numerically by Hyon et al., in an effort which included comparison to new experiments using {\it B. subtilis} cells \cite{hpsf12}, and precession was also noted in simulations by Shum \& Gaffney \cite{sg12}. More recently, Constantino et al. have observed and rationalized the helical trajectories of {\it H. pylori} \cite{cjfb18}, Rossi et al. have investigated the same for {\it E. gracila} cells \cite{rcbnd17,grnd21}, synthetic models have been designed to explore helical trajectories \cite{tt18}, and new techniques have been developed for inferring motility parameters statistically \cite{cmjdpb19}. Properly tuned undulatory beating can also result in helical navigation, as found in the swimming of {\it Chlamydomonas reinhardtii} cells \cite{cw21}.

Parallel to any potential functionality which might be conferred by hook deformability, there are also constraints and requirements on its construction for usability. Vogel \& Stark have presented a very detailed picture of flagellum buckling and dynamics, and a supercritical Hopf bifurcation in the thrust force due to hook and flagellum compliance which persists when the flagellum is affixed to a spherical cell body during locomotion \cite{vs12}. Full numerical simulations and modeling by Shum \& Gaffney \cite{sg12}, Nguyen \& Graham \cite{ng17,ng18}, \tcb{and Park et al. \cite{pkl19}} confirmed the existence of a critical motor torque triggering a bifurcation from straight swimming to apparently helical swimming trajectories when flagellum flexibility is included (see also Refs.~\cite{sg12,sgs13,tt18}). 

%More detailed discussions which include hydrodynamic forces on helical flagella include early works by Keller \& Rubinow \cite{kr76} (see also \cite{Purcell77}). 
%From Edelstein-Keshet, need to investigate: (Brokaw, 1958a; Chwang and Wu, 1971; Keller and Rubinow, 1976; Naitoh and Sugino, 1984; Fenchel and Jonsson, 1988; Sugino and Naitoh, 1988; Crenshaw, 1989). Furthermore, many works are done in understanding the efficiency of the propulsion of those microorganisms \cite{sl11, r18,k14,Higdon79,cmyw06,c71}.

In this paper we show using a simplified model of the flagellum and the flagellar hook that a helical trajectory is generic, and study the dependence of the helical path geometry on material parameters including the flagellum geometry, hook bending stiffness, and spontaneous hook curvature. After providing the mathematical description of the swimming cell in \S II, we reproduce in \S III by very elementary means a modification of the result of Crenshaw \cite{c931} that a helical trajectory is generic, through reference to the curvature and torsion observed in a particular reference frame. In \S IV we develop the simplified model of a phase-averaged helical flagellum connected by a flexible hook to the cell body which experiences a fixed motor torque. The model is then used to derive analytically the flagellum dynamics when the cell body is fixed in space, and then in the full system in which the cell body, flagellar motion, and hook angle are coupled. We provide analytical expressions for the critical stability criteria, a bifurcation in the hook bending angle which is present in the full swimming problem but not in the fixed-body problem given the assumption of a rigid flagellum. We conclude with a brief discussion in \S V.

\section{Mathematical model} \label{sec:model}

\begin{figure}[htbp]
\begin{center}
\includegraphics[width=.7\textwidth]{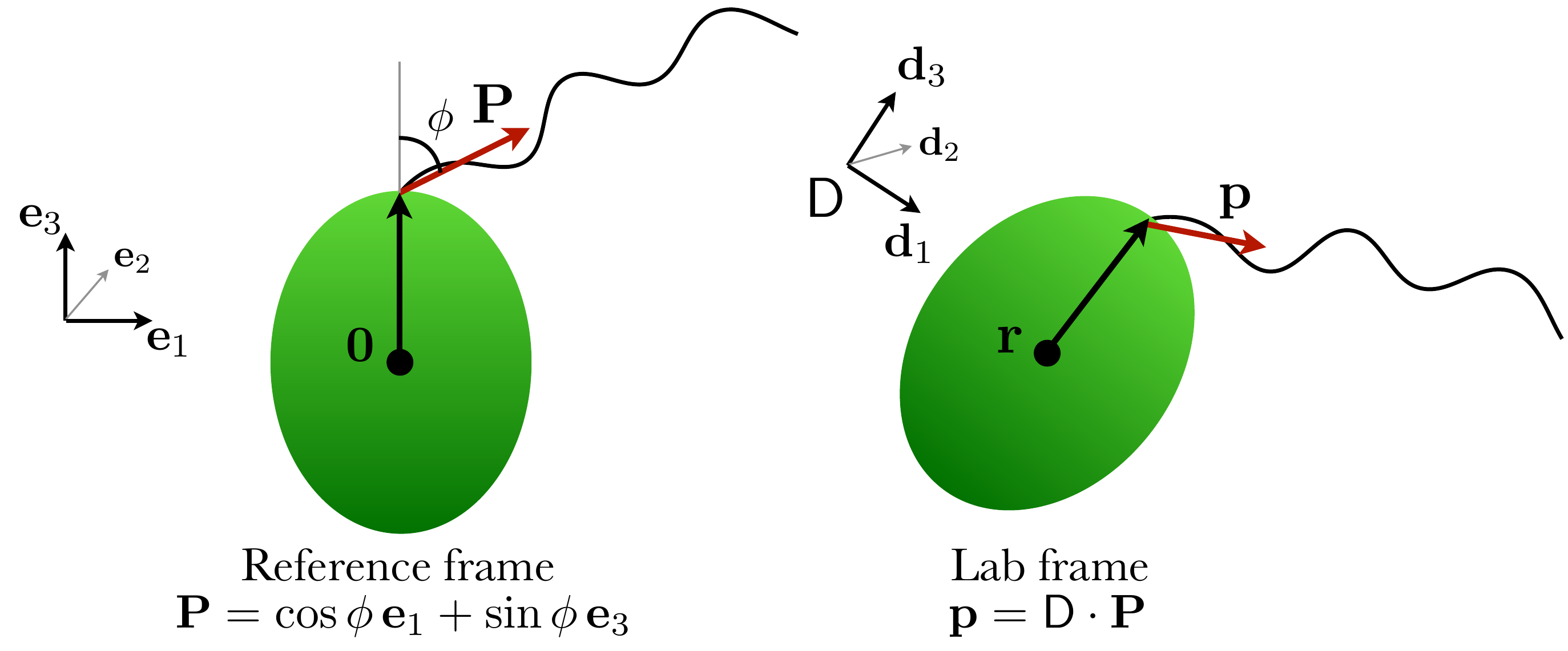}
\caption{Schematic of the prolate spheroidal cell body (axis lengths $2a$ and $2a\sqrt{1-e^2}$, with $e\in[0,1)$ the eccentricity) and flagellum in the reference and lab frames. (Left) Reference frame. The flagellum is affixed to a hook emanating from the position $a\b{e}_3$, and is oriented in the $\b{P}(t)$ direction. (Right) Lab frame. The centroid is located at a position $\b{r}(t)$, the hook emanates from a position $\b{x}(t)=\b{r}(t)+a\b{d}_3$, and the flagellum is oriented in the $\b{p}(t)$ direction which is always in the $(\b{d}_1,\b{d}_3)$ plane. Mapping from the reference frame to the lab frame involves a translation by $\b{r}(t)$ and a rotation via matrix $\D(t)$. The bending angle is denoted by $\phi(t)$.}
\label{Schematic1}
\end{center}
\end{figure}

In this paper the cell body is modeled as a prolate spheroid with major and minor axis lengths $2a$ and $2b$, whose position and orientation at time $t$ evolve according to hydrodynamic force and torque balance. Driving the body through the fluid is a single flagellum connected to the cell by a short, deformable hook. As illustrated in Fig.~\ref{Schematic1}, we define the reference frame to be that in which the centroid of the cell body is located at the origin, and the flagellum orientation, identified by a unit vector $\b{P}$, lies in the $(\b{e}_1,\b{e}_3)$-plane, with $\{\b{e}_1,\b{e}_2,\b{e}_3\}$ the standard orthonormal basis. We write $\b{P}=\sin\phi\b{e}_1+\cos\phi\b{e}_3$, and refer to the angle $\phi$ as the bending angle. The rotary motor which drives the relative rotation between the cell body and flagellar hook is located in the reference frame at position $\b{X}=a\b{e}_3$.

In the lab frame, the centroid is located at time $t$ at a position $\b{r}(t)$, and the flagellum orientation is denoted by $\b{p}(t)$. The centroid of the cell body evolves as $\dot{\b{r}} = \b{U}(t)$, with $\b{U}$ the translational velocity \tcb{in the lab frame}. The cell's rotational velocity \tcb{in the lab frame} is denoted by $\O$. As the system translates and rotates it carries with it an orthonormal triad of basis vectors, the columns of the rotation matrix $\D (t) =\{\b{d}_1(t),\b{d}_2(t), \b{d}_3(t)\}$, defined such that the position of the flagellar base in the lab frame is given by $\b{x}(t) = \b{r}(t)+a \b{d}_3(t)$, and the flagellum direction remains always in the $(\b{d}_1,\b{d}_3)$-plane ($\b{P}\in \mbox{span}\{\b{e}_1,\b{e}_3\}$ and $\b{p}\in \mbox{span}\{\b{d}_1,\b{d}_3\}$\tcb{, with $\b{p}=\D\cdot \b{P}$}). The rotational velocity of the flagellum's {\it orientation} in the lab frame is denoted by $\O^p$, with $\O^p=\O+\dot{\phi} \b{d}_2+\dot{\theta} \b{d}_3$, with $\dot{\theta}$ a flagellar precession rate. Hence the rotational velocity of the orthonormal basis, denoted by $\O_D$, is given by (with $\b{d}_3\b{d}_3$ a dyadic product),
\begin{gather}
\O_D = \left(\I-\b{d}_3\b{d}_3\right)\cdot\O+\b{d}_3\b{d}_3\cdot  \O^p = \O + \dot{\theta}\b{d}_3.
\end{gather}
\tcb{Denoting the} translational and rotational velocities in the reference frame by $\b{U}_0$ and $\Oo$, respectively, we have $\b{U} = \D \cdot \b{U}_0$ and $\O = \D \cdot \Oo$, and thus $\O_D = \D \cdot (\Oo+\dot{\theta}\b{e}_3)$. Since the basis vectors (columns of $\D$) evolve in time via
\begin{gather}
\b{\dot{d_i}} = \O_D \times \b{d_i}=(\D\cdot\Oo) \times \b{d_i}+\dot{\theta}\b{d}_3\times \b{d}_i,
\end{gather}
the rotation matrix evolves according to (using $\det(\D)=1$),
\begin{gather}
\dot{\D} = \o_D\cdot \D = (\D\cdot \o_0\cdot \D^T+\o' )\cdot \D,
\end{gather}
where we have introduced the skew-symmetric operators $\o_D=\O_D \times$, $\o_0=\O_0 \times$, and $\o'=\dot{\theta}\b{d}_3\times=\dot{\theta}(\b{d}_2\b{d}_1-\b{d}_1\b{d}_2)$.
%\begin{gather}
%\o_D  = \left(\begin{array}{ccc}0 & -\Omega_3 & \Omega_2 \\ \Omega_3 & 0 & -\Omega_1 \\ -\Omega_2 & \Omega_1 & 0 \end{array}\right),\,\,\,\,\,\,\,\o' = \dot{\theta}(\b{d}_2\b{d}_1-\b{d}_1\b{d}_2),
%\end{gather}

\tcb{Shortly we will use the following relations:} for a constant vector $\b{a}$ \tcb{in the reference frame}, 
\begin{gather}
\frac{d}{dt}\left(\D\cdot \b{a}\right)=\dot{\D}\cdot \b{a} = (\D \cdot \o_0\cdot \D^T+\o') \cdot (\D \cdot \b{a}) =\D \cdot \left([\O_0+\dot{\theta}\b{e}_3] \times \b{a}\right).
\label{vec_identity}
\end{gather}
%\tcb{A third frame of interest is one in which the body is fixed, with the far-field appearing to translate and rotate with velocities $-\Uo$ and $-\Oo$, and the flagellum rotating with velocity} denoted by $\Oo^P$. \tcb{In the reference frame the body may spin about the $\b{e}_3$ direction; in this (body) frame it is the flagellum which may precess about the $\b{e}_3$ direction.} 
\tcb{Then, writing the flagellum orientation's rotational velocity in the lab frame as $\O^p=\O+\dot{\theta}\b{d}_3+\dot{\phi}\b{d}_2=\D\cdot\left(\Oo+\dot{\theta}\b{e}_3+\dot{\phi}\b{e}_2\right)$ and using $\dot{\b{P}}=\left(\cos\phi \,\b{e}_1-\sin\phi\,\b{e}_3\right)\dot{\phi}=\dot{\phi}\b{e}_2\times \b{P}$ \tcb{(see also Appendix A)}, we verify that}
\begin{gather}
\dot{\b{p}}=\frac{d}{dt}\left(\D\cdot \b{P}\right) = \dot{\D}\cdot \b{P}+\D \cdot \dot{\b{P}} = \D \cdot \left[\left(\O_0+\dot{\theta}\b{e}_3+\dot{\phi}\b{e}_2\right) \times \b{P}\right]=\O^p\times \b{p}.
\end{gather}

\section{Constant bending angle results in a helical trajectory}\label{sec:trajec}

We begin by reproducing a modification of the result of Crenshaw \cite{c931}, that a helical trajectory is almost entirely generic at zero Reynolds number. While Crenshaw shows that a constant translational and rotational velocity in the body frame result in a helical trajectory, here we lean on the axisymmetry of the cell body and produce a similar result by using a reference frame which rotates relative to the flagellar orientation. The primary assumptions that we make are that the flagellum geometry is rigid and that the force and torque generated by the rotation of the flagellum are independent of its phase (thus neglecting small oscillations due to the rotation of a finite helical propeller \cite{kr76}). It is believed that flagella do not substantially deform during their normal rotating thrust generation \cite{dtrb07}.

The translation and rotation rates of the cell body, as well as the evolution of the flagellar orientation, are determined instantaneously in the Stokes flow limit of zero Reynolds number \cite{Childress81}. Since the flagellum orientation $\b{P}$ in the reference frame is described by a single degree of freedom $\phi\in[0,\pi]$, \tcb{and $\dot{\phi}$ is a function of $\phi$ alone by the assumption of phase-independence and body axisymmetry,} we must have that $\phi$ settles to a fixed point (neglecting the special case that the flagellum is in periodically varying hard contact with the cell body). Viewed differently, in the frame of the cell body the flagellum orientation vector explores the two-dimensional surface of a sphere, in which case the autonomous system demands that the orientation settles either to a fixed point or a limit cycle. A fixed point is found if either $\phi=0$ (or the nonphysical orientation with $\phi=\pi$), or when the relative precession rate between the flagellum and cell body, $\dot{\theta}$, is zero. Otherwise, a limit cycle must exist by the Poincar\'{e}-Bendixson theorem \cite{Meiss07}.

We begin by considering the terminal case, in which the flagellum orientation has reached its fixed point in the reference frame (i.e. the bending angle $\phi$ is constant). By assumption of flagellar phase independence, the cell body velocity and rotation rate in the reference frame, $\Uo $ and $\Oo$, are also constant in time, as is the differential precession rate, $\dot{\theta}$. We will prove that the resulting trajectory is helical by showing that its curvature, $\kappa$, and torsion, $\tau$, are constant. In terms of the position of the centroid $\b{r}(t)$, and $\dot{\b{r}}\equiv d\b{r}/dt$, the curvature and torsion are given by
\begin{gather}
\kappa =\frac{|\rp \times \rpp|}{|\rp|^3},\,\,\,\, \tau =\frac{-(\rp  \times \rpp)\cdot \rppp}{|\rp \times \rpp|^2}\label{tau}.
%\kappa =\frac{|\b{r}\,' \times \b{r}\,''|}{|\b{r}\,'|^3}, \label{eq:1}  \\ 
%\tau =\frac{-(\b{r}\,' \times \b{r}\,'')\cdot \b{r}\,'''}{|\b{r}\,' \times \b{r}\,''|^2}. \label{eq:2}
\end{gather}
With velocities in the reference frame fixed, we have that 
\begin{gather}
\rp= \D  \cdot \Uo,\\
\rpp = \dot{\D}\cdot \Uo  =\D\cdot\left([\Oo+\dot{\theta}\b{e}_3]  \times \Uo \right).
\end{gather}
Upon insertion into \eqref{tau}, we find using manipulations as in \eqref{vec_identity}, and $|\rp|=|\Uo |$ that
\begin{gather}\label{eq:kappaderiv}
\rp \times \rpp = \D \cdot \left(\Uo  \times ([\Oo+\dot{\theta}\b{e}_3]  \times \Uo )\right),\\
\kappa = \frac{\Big|\D\cdot \left(\Uo  \times ([\Oo+\dot{\theta}\b{e}_3] \times \Uo)\right)\Big| }{|\D\cdot \b{U}_0|^3}=\frac{\Big|\Uo  \times \left([\Oo+\dot{\theta}\b{e}_3] \times \Uo\right)\Big|}{|\b{U}_0|^3},
\end{gather}
a constant in time by the assumptions above. The curvature may also be written as
\begin{gather}
\kappa =\frac{\sqrt{A_1 A_2-A_3^2}}{A_1},
\end{gather}
where we have defined
\begin{gather}\label{ABQ}
A_1 = |\Uo|^2, \,\,\,\, A_2=|\Oo+\dot{\theta}\b{e}_3|^2,\,\,\,\, A_3= \Uo\cdot (\Oo+\dot{\theta}\b{e}_3).
\end{gather}
Similarly, and again using the identity in \eqref{vec_identity}, we have (again assuming $\dot{\theta}$ is constant) that
\begin{gather}
\rppp = \dot{\D}\cdot([\Oo+\dot{\theta}\b{e}_3]  \times \Uo )
 = \D \cdot  \left(\left(\Oo+\dot{\theta}\b{e}_3\right) \times ([\Oo+\dot{\theta}\b{e}_3]  \times \Uo )\right),
%(\o \cdot \D \cdot \Oo ) \times \D \cdot \Uo  + \D \cdot \Oo  \times (\o \cdot \D \cdot \Uo  )= \\
%(\D \cdot \Oo  \times \D \cdot \Oo ) \times \D \cdot \Uo + \D \cdot \Oo  \times (\D \cdot \Oo  \times \D \cdot \Uo )= \\
%\D \cdot \Oo  \times [\D \cdot (\Oo  \times \Uo )]= 
\end{gather}
and therefore
\begin{gather}
\tau = \frac{-\D \cdot \left(\Uo  \times \left([\Oo+\dot{\theta}\b{e}_3]  \times \Uo \right)\right)\cdot \left\{\D \cdot \left([\Oo+\dot{\theta}\b{e}_3]  \times ([\Oo+\dot{\theta}\b{e}_3]  \times \Uo )\right)\right\}}{|\D \cdot [\Uo  \times ([\Oo+\dot{\theta}\b{e}_3]  \times \Uo )]|^2}\\
=\frac{-\left(\Uo  \times \left([\Oo+\dot{\theta}\b{e}_3]  \times \Uo \right)\right)\cdot \left\{\left(\Oo+\dot{\theta}\b{e}_3\right)  \times \left([\Oo+\dot{\theta}\b{e}_3]  \times \Uo \right)\right\}}{|  \Uo  \times ([\Oo+\dot{\theta}\b{e}_3]  \times \Uo )|^2}=\frac{-A_3}{A_1},
\end{gather}
constant in time. The trajectory is therefore helical with curvature and torsion given by
\begin{gather}\label{eq: kappa-tau}
\kappa =\frac{\sqrt{A_1 A_2-A_3^2}}{A_1},\,\,\,\,\,\,\tau=\frac{-A_3}{A_1}.
\end{gather}

The helical trajectory may alternatively be represented by its amplitude \tcb{$A$, with $A=\kappa/(\kappa^2+\tau^2)$} and its slope $|\tau|/\kappa$. The pitch angle \tcb{$\psi$} is given by $\psi=\tan^{-1}(\kappa/|\tau|)$ ($\psi=0$ for a straight path and $\psi=\pi/2$ for a tightly coiled helical path). The helical amplitude and pitch angle are thus given by $A=(A_1A_2-A_3^2)^{1/2}/A_2$, and $\psi=\tan^{-1}\left(\sqrt{A_1A_2-A_3^2}/|A_3|\right)$.

The result above is highly generic. Even if the complete hydrodynamics of the flagellum-cell system are determined exactly, once the (assumed rigid and phase-insensitive) flagellum reaches the fixed point in bending angle the trajectory is assured to be helical, while the flagellum precesses around the normal direction and the cell body translates and rotates, per the calculation above. The terminal flagellum orientation and thus geometry of that helical trajectory may, however, depend on the detailed hydrodynamics of flagellar propulsion. The finite length of real flagella will also contribute very small oscillations due to a small hydrodynamic phase dependence, but the time required for flagellum rotation about its long axis is often small compared to all other timescales in the system, and there may be considerable cancellation of these effects in more general settings.

\section{Trajectories of cell bodies with a model helical flagellum}\label{sec: Traj}

\subsection{Dynamics of an axisymmetric body}
We turn now to the full dynamics of an axisymmetric cell body in a viscous fluid. The linearity of the Stokes equations describing viscous flow demand a linear relationship between the force and torque due to the flagellum and the resultant translational and rotational velocities of the cell body \cite{hb65}. \tcb{The net force and torque on the full body-flagellum system must be zero at any moment in Stokes flow, resulting in cell body counter-rotation relative to the flagellum \cite{Purcell77}.} Although the result of the previous section is generic, we will neglect the hydrodynamic interactions between the cell body and the propelling mechanism for the sake of analytical tractability; \tcb{an approximation by Lighthill, (eqns. 124-125) in Ref.~\cite{Lighthill76}, indicate a correction of the swimming speed which is only logarithmic in the cell body size relative to flagellum length, with noticeable but not dramatic effects noted in Refs.~\cite{Higdon79,ptr87,cw09,tmasw15,pkl19}}. Writing the viscous force and torque on the cell body in the reference frame at its centroid as $\b{F}_0(\b{P})$ and $\b{M}_0(\b{P})$, the viscous resistance of a prolate spheroidal cell body results in the translational and rotational velocities in the reference frame,
\begin{gather}
\Uo = \frac{-1}{6\pi\mu a} \left[(X^A)^{-1}\b{e}_3\b{e}_3+(Y^A)^{-1}\left(\b{I}-\b{e}_3\b{e}_3\right) \right]\cdot \b{F}_0(\b{P}) \label{eqU0},\\
\Oo =\frac{-1}{8\pi\mu a^3} \left[(X^C)^{-1}\b{e}_3\b{e}_3+(Y^C)^{-1}\left(\b{I}-\b{e}_3\b{e}_3\right) \right]\cdot \b{M}_0(\b{P}),\label{eqO0}
\end{gather}
where, defining the body eccentricity $e$ and $L_e=\ln[(1+e)/(1-e)]$, we have $X^A=(8 e^3/3)[-2e+(1+e^2)L_e]^{-1}$, $Y^A=(16 e^3/3)[2e+(3e^2-1)L_e]^{-1}$, $X^C=(4 e^3/3)(1-e^2)[2e-(1-e^2)L_e]^{-1}$, and $Y^C=(4 e^3/3)(2-e^2)[-2e+(1+e^2)L_e]^{-1}$ \cite{kk91}. If the cell body is spherical, $e \to 0$, $X^A=Y^A=X^C=Y^C=1$, and
\begin{gather}
\Uo = \frac{-1}{6\pi\mu a} \b{F}_0(\b{P}) ,\\
\Oo =\frac{-1}{8\pi\mu a^3}\b{M}_0(\b{P}).
\end{gather}
To evaluate the swimming path geometry in Eq.~\eqref{eq: kappa-tau} we require $\b{F}_0$ and $\b{M}_0$ and the flagellum precession rate, $\dot{\theta}$, which all require a model of the flagellum, which we develop in the following section. For a concrete example, however, consider a flagellum connected at $\b{X}=a\b{e}_3$, oriented in the $\b{P}=\sin\phi\,\b{e}_1+\cos\phi \,\b{e}_3$ direction, with the bending angle $\phi$ and precession rate $\dot{\theta}$ held fixed. If the flagellum is acting on the cell with force $-f\b{P}$ and moment $-m \b{P}$, then $\b{F}_0=f\b{P}$ and $\b{M}_0=a\b{e}_3\times \b{F}_0+m\b{P}$, and for nearly spherical bodies we find that
\begin{gather}\label{synthetichelix}
\kappa =\frac{3}{4 a} \left(1+\left[\frac{8\pi \mu a^2\dot{\theta}}{f}\right]^2\right)^{1/2} \sin (\phi )+O(e^2),\\
\tau =\frac{3}{4 a^2 f} \left(8 \pi \mu  a^3  \cos (\phi ) \dot{\theta}-m\right)+O(e^2),
\end{gather}
as $e \to 0$. The trajectory is insensitive to the cell body asphericity to first order in the body eccentricity. 

%The helical trajectory of the body centroid can be written in general coordinates by decomposing $\Uo$ into components parallel and orthogonal to $\Oo$, resulting in
%\begin{gather}
%\b{r}(t) = \b{r}(0)+ \frac{ Q }{B}\Oo t + \frac{1}{B^{1/2}}\left(\Uo - \frac{Q }{B}\Oo\right)\sin(B^{1/2}t) - \frac{1}{B}\left(\Oo\times \Uo \right) \cos(B^{1/2}t).
%\end{gather}
%
%The amplitude of the helical trajectory is 
%\begin{gather}
%\frac{1}{B^{1/2}}\Big|\Uo - \frac{Q}{B}\Oo\Big| =\frac{4 a^3 f^2 \sin\theta}{3 \left(a^2 f^2 \sin ^2\theta+\ell^2\right)},
%\end{gather}
%the pitch is
%\begin{gather}
%2\pi\frac{Q}{B} = \frac{8 \pi  a^2 f \ell}{3 \left(a^2 f^2 \sin ^2\theta+\ell^2\right)},
%\end{gather}
%and the pitch angle is
%\begin{gather}
%\psi = \tan^{-1}\left(\frac{a f \sin\theta}{\ell}\right).
%\end{gather}
%If $|\ell| \ll a|f|$, then $A\approx 4a/(3 \sin\theta)$, $\lambda\approx 8 \pi  \ell/(3 f \sin^2\theta)$, and $\psi =\pi/2-\ell/(a f \sin\theta)$.

\subsection{Model flagellum}\label{sec: model flagellum}

In this section we will derive the translational and rotational velocities of the cell body, as well as the precession rate, $\dot{\theta}$, as functions of the flagellar orientation in the reference frame, $\b{P}$. In addition we seek an equation for the time evolution of the bending angle, $\phi$. The model of the flagellum will incorporate its geometry and the nontrivial relationship between the motor torque and its dynamics via the flexible hook, to second order in the flagellum amplitude. For this purpose we use the simplest resistive force theory approximation \cite{gh55,jb79} as recently used in similar contexts, including the instability of bodies propelled by $N$ flagella or swimming with a flexible flagellum near a wall \cite{il19,dl19,Ishimoto19}, and neglecting hydrodynamic interactions with the cell body. Comparisons between this resistive force theory and full hydrodynamic theory have been explored in detail \cite{jb79,rcslz13}; a comparison for this precise context in Ref.~\cite{ng18} suggests sufficient accuracy of the simpler resistive force theory for our purposes. 

The flagellum is modeled as a slender left-handed helical filament of length $L$ and amplitude $b$, with aspect ratio (diameter/length, ``slenderness'') denoted by $\e$, with $0 < \e \ll 1$. We parameterize the flagellum by arc-length $s\in[0,L]$, describing the position on the flagellum at station $s$ and time $t$ in the reference frame by $\b{X}^f(s,t)=\b{X}+\alpha s \b{P}(t)+b \left(\cos(ks-\delta)\b{P}^\perp(t)-\sin(ks-\delta)\b{P}^{\perp \perp}(t)\right)$, with $\{\b{P},\b{P}^\perp,\b{P}^{\perp\perp}\}$ an orthonormal basis. The position $\b{X}=a\b{e}_3$ denotes the constant location in the reference frame of the rotary motor connecting the cell body to the flagellar hook. We set $\alpha =\sqrt{1-(k b)^2}$ so that the parameter $s$ is the arc-length, and $\delta$ is a phase constant. The tapering of the helical radius needed to bridge the hook to the flagellum is neglected for convenience; the inclusion of tapering has been found elsewhere to play a minimal role in this context \cite{sg12}. Assuming the flagellum to be rigid with translation velocity $\b{U}_0^f$ and rotation rate $\Oo^f$ in the reference frame, the filament velocity (and thus fluid velocity, by the assumption of a no-slip boundary condition) at station $s$ along the filament is given in the reference frame by $\b{u}_0(s) = \b{U}_0^f +\Oo^f \times \left(\b{X}^f(s,t)-\b{X}\right)$. 

%To relate back to \S\ref{sec:model} and Fig.~\ref{Schematic1}, the rotational velocity of the flagellum in the {\it body} frame, $\Oo^f-\Oo$, is composed of the rotation rate of its orientation vector $\b{P}$ along with an additional (unknown) spin rate $\omega$ about $\b{P}$, and we write $\Oo^f -\Oo = \omega \b{P}+\Oo^P=\omega \b{P}+\dot{\phi}\b{e}_2+\dot{\theta}\b{e}_3$. The rate of change of the bending angle, the precession rate, and the spin rate are then computed as
%\begin{gather}
%\dot{\phi}=\b{e}_2\cdot(\Oo^f-\Oo),\label{dphi}\\
%\dot{\theta}=\left(\b{e}_3-\cot(\phi)\b{e}_1\right)\cdot (\Oo^f-\Oo),\\
%\omega = \csc(\phi)\b{e}_1 \cdot (\Oo^f-\Oo).\label{do}
%\end{gather}

The rotational velocity of the flagellum in the reference frame is composed of the body rotation rate, the rotation rate of its orientation vector $\b{P}$, the (unknown) precession rate $\dot{\theta}$ about $\b{e}_3$, and finally the spin rate $\omega$ about $\b{P}$, or $\Oo^f =\Oo^p+ \omega \b{P}=\Oo+\dot{\phi}\b{e}_2+\dot{\theta}\b{e}_3+\omega \b{P}$. The rate of change of the bending angle, the precession rate, and the spin rate are then computed via the relative rotation rate between the flagellum and the cell body,
\begin{gather}
\dot{\phi}=\b{e}_2\cdot(\Oo^f-\Oo),\label{dphi}\\
\dot{\theta}=\left(\b{e}_3-\cot(\phi)\b{e}_1\right)\cdot (\Oo^f-\Oo),\\
\omega = \csc(\phi)\b{e}_1 \cdot (\Oo^f-\Oo).\label{do}
\end{gather}

The reference frame viscous drag per unit length on the flagellum, $\b{f}_0$, we model using resistive force theory,
\begin{gather}
\b{f}_0(s) = -\frac{8\pi\mu}{c+2}\left[\I + \frac{2-c}{2c}\b{\b{\hat{s}}\b{\hat{s}}}\right]\cdot \b{u}_0(s),
\end{gather}
where $c=\log(1/\e^2)-1$ $(>0)$, and $\b{\hat{s}} = \partial_s \b{X}^f(s)$ is the unit tangent vector.
%\begin{gather}
%8\pi \mu \b{u}(s) = -[(c+2)\b{I}+(c-2)\b{\b{\hat{s}}\b{\hat{s}}}]\cdot \b{f}(s),
%\end{gather}
The total viscous force and torque on the flagellum measured relative to $\b{X}$, denoted by $\b{F}_0^f$ and $\b{M}_0^f$, respectively, are determined by integration over the filament length,
\begin{gather}
\b{F}_0^{f}=\int_0^L \b{f}_0(s) \,ds,\\
\b{M}_0^f=\int_0^L (\b{X}^f(s)-\b{X})\times \b{f}_0(s) \,ds.
\end{gather}
Averaging over the phase, $\delta$, and inserting $\b{f}$ from above, we have the general linear resistance relations
\begin{gather}
\b{F}_0^f=-\Atens\cdot \b{U}_0^f-\Btens\cdot \Oo^f,\label{resistanceF}\\
\b{M}_0^f=-\Btens^T\cdot \b{U}_0^f-\Dtens\cdot \Oo^f\label{resistanceL},
\end{gather}
(see Ref.~\cite{hb65}) where the individual block operators are simplified using (phase-averaged) axisymmetry, 
\begin{gather}
\Atens =\left(\frac{2\pi\mu L}{c}\right)\left(\xi_1 \b{PP} + \xi_2 \left(\b{I}-\b{PP}\right)\right),\\
\Btens =\left(\frac{2\pi\mu L^2}{c}\right)\left(\eta_1 \b{PP}  + \eta_2 \left(\b{I}-\b{PP}\right)+ \eta_3\left(\b{P}^\perp \b{P}^{\perp\perp}-\b{P}^{\perp\perp} \b{P}^{\perp}\right)\right),\\
\Dtens =\left(\frac{2\pi\mu L^3}{c}\right)\left(\zeta_1 \b{PP} + \zeta_2 \left(\b{I}-\b{PP}\right)\right).
\end{gather}
Keeping terms up to $O(b^4k^4)$ in a small-amplitude approximation of the flagellum, we compute the dimensionless quantities
\begin{gather}
\xi_1= 2(1+ b ^2 k^2),\,\,\,\, \xi_2= 4-b ^2 k^2,\\
\eta_1= 2 b ^2 k/L,\,\,\,\, \eta_2=-b ^2 k/L,\,\,\,\, \eta_3= 2-3 b ^2 k^2/2,\\
\zeta_1= 4 b ^2/L^2,\,\,\,\,\zeta_2= \tfrac13\left(4-5 b ^2 k^2+3b ^2/L^2\right).
\end{gather}

The off-diagonal tensor may be made symmetric by a parallel translation to the center of hydrodynamic reaction, but since we will match forces and torques at an endpoint we leave the structure above. 

Inverting the resistance equations \eqref{resistanceF}-\eqref{resistanceL}, we find the mobility relations,
\begin{gather}
\b{U}^f_0=-\tilde{\Atens}\cdot \b{F}^f_0-\tilde{\Btens}\cdot \b{M}_0^f,\label{mobilityF}\\
\O^f_0=-\tilde{\Btens}^T\cdot \b{F}^f_0-\tilde{\Dtens}\cdot \b{M}_0^f,\label{mobilityL}
\end{gather}
where the individual block operators are given by
\begin{gather}
\tilde{\Atens} =\left(\frac{c}{8 \pi  \mu  L}\right)\left(\alpha_1 \b{PP} + \alpha_2 \left(\b{I}-\b{PP}\right)\right),\\
\tilde{\Btens} =\left(\frac{c}{8 \pi  \mu  L^2}\right)\left(\beta_1 \b{PP}+\beta_2 (\mathsf{I}-\b{PP})+\beta_3\left(\b{P}^\perp\b{P}^{\perp\perp}-\b{P}^{\perp\perp}\b{P}^{\perp}\right)\right),\\
\tilde{\Dtens} =\left(\frac{c}{8 \pi  \mu  L^3}\right)\left(\gamma_1\b{PP}+\gamma_2\left(\mathsf{I}-\b{PP}\right)\right),
\end{gather}

with the dimensionless quantities 
\begin{gather}
\alpha_1=2-b ^2 k^2,\,\,\,\, \alpha_2= 4+b^2k^2-9b^2/L^2,\\
\beta_1= -k L\left(1-b^2 k^2/2\right),\,\,\,\, \beta_2= 3 b^2 k/L,\,\,\,\, \beta_3= -\tfrac{3}{2}\left(4+3b^2k^2-12b^2/L^2\right),\\
\gamma_1= \frac{L^2}{4b^2} \left(4+2 b^2 k^2-b^4k^4\right),\,\,\,\, \gamma_2= 3\left(4+5 b^2 k^2- 12b^2/L^2\right).
\end{gather}

The viscous force and torque on the flagellum, $\b{F}_0^f$ and $\b{M}_0^f$, depend on the interaction with the cell body as mediated by the flexible hook. In the following sections we consider the cases of: prescribed forces and moments acting on the flagellum; a flexible connection to a stationary cell body; and a flexible connection to a freely moving cell body. 

\subsection{No cell body, prescribed forces and moments}

\tcb{Although the force and moment are not generically aligned with $\b{P}$, we consider as a simple example the case that} the flagellum is subject to a force $f\b{P}$ and a moment $m\b{P}$ applied at the basal connection point, $\b{X}$. Then the viscous force and torque on the flagellum about the basal connection point are $\b{F}_0^f=-f\b{P}$ and $\b{M}_0^f = -m \b{P}$, and the resulting velocity and rotation rate about $\b{X}$ are given by $\b{U}^f_0 = U_f \b{P}$ and $\b{\Omega}^f_0 = \Omega_f \b{P}$, where \tcb{to leading order in $c$, assumed large,}
\begin{gather}\label{mobility2}
U^f=\frac{c}{8 \pi \mu L} \left(2-b^2 k^2\right) \left(f-\frac{k }{2}m\right)+O(b^4k^4),\\
\Omega^f=\frac{c k}{16\pi \mu L}\left( -\left(2-b ^2 k^2\right)f+k\left(\frac{2}{b^2k^2}+1-\frac{b ^2k^2}{2}\right)m\right)+O(b^4k^4).
\end{gather}
If $f=0$ and $m>0$ then the left-handed helix rotates about the $\b{P}$ axis ($\Omega^f>0$), with waves appearing to pass in the direction away from the cell and generating a force on the fluid in the $\b{P}$ direction; the fluid reaction pushes the cell in the $-\b{P}$ direction ($U^f<0$). These directions are flipped for a right-handed helix ($k \to -k$). 

A subtle point arises in the limit that $b \to 0$, in which the helix tends towards the shape of a straight rod. Since we have neglected the rotational drag around the long axis of the filament locally, which scales as $O(\e^2)$ as $\e \to 0$, the rotation rate above tends towards infinity as $b \to 0$. Nevertheless the coupling between the torque and the translational speed of such a filament does not vanish in this limit, instead it tends towards dependence on the wavenumber, $k$. A more detailed examination of this asymptotic regime shows that the model above amounts to a distinguished limit, accurate so long as $\e L/b$ is held fixed as $b \to 0$ \cite{ls14}. At leading order in $c$, the tensors above remain accurate; contributions which depend on the fixed value of $\e L/b$ enter at $O(1)$. We refer the reader also to the related model by Vogel \& Stark in Ref.~\cite{vs12}, there named the `helical rod model', which in addition to hydrodynamics incorporates filament deformability and bending stiffness. 

\subsection{Model hook, and a fixed cell body}\label{section: fixed cell}

The relationship between the viscous force and torque on the cell body and those on the flagellum are mediated by a small flexible hook, which we model using a discretization of the Kirchhoff elastic rod theory with only a single joint. The effective (integrated) curvature is related to the bending angle $\phi$ with the elastic energy stored in the hook (i.e. in the joint connecting the cell body to the flagellum); penalizing deviations of the curvature away from a preferred angle, $\phi_0$, we write $\mathcal{E}_b=(2B/\ell_h)\left(\tan\left(\phi/2\right)-\tan\left(\phi_0/2\right)\right)^2$, where $B$ is the bending stiffness (with units of energy times length) and $\ell_h$ is the length of the hook \cite{bwrag08,OReilly17}. This energy penalty results in the Bernoulli-Euler elastic moment internal to the hook in the lab frame given by $(2B/\ell_h)\left(\tan\left(\phi/2\right)-\tan\left(\phi_0/2\right)\right)\b{d}_2$. Experimental measurements report $\ell_h\sim 50-100$nm, which is much smaller than standard flagellum lengths $L\sim 1-10\mu$m \cite{bbb89,bbb91,mmfd93,hyoa94,kssogsa03,sng04,ktkh10,sgs13}.  

The base of the hook is assumed to be oriented normal to the surface at its connection point to the cell body. The short hook (and thus flagellum) is driven at its base with a constant motor torque $M_a \b{e}_3$, with $M_a$ assumed known. This moment must be balanced by a viscous torque on the cell body, or $\b{e}_3\cdot \b{M}_0=M_a$. More detailed continuous Kirchhoff rod models have been used in related numerical studies \cite{vs12,ng17,jf18,pkl19,jf20}; see also Ref.~\cite{OReilly17}.

%As an approximation expected to be most accurate for long slender flagella we take the rotation to be transmitted perfectly to the flagellum along its long axis only, $\b{P} \cdot \Oo^f = \b{e}_3\cdot \Oo^h$, so that $\omega =\b{P}\cdot \Oo^f- \b{e}_3\cdot \Oo$. A Lagrange multiplier, $M_\tau$, representing an additional internal moment is included to enforce this kinematic constraint. Specifically with the mean orientation of the model hook written as $(\b{e}_3+\b{P})/|\b{e}_3+\b{P}|$, the additional moment is $M_\tau(\b{e}_3+\b{P})/|\b{e}_3+\b{P}|$. 

With the hook assumed to be very short compared to other length scales in the problem, we arrive at the following system of equations:
\begin{gather}
\b{F}_0+\b{F}_0^f=\b{0},\label{fullsys1}\\
\b{M}_0 +a\b{e}_3\times \b{F}_0^f+ \b{M}^f_0=\b{0},\\
\b{M}^f_0 -2B(\tan(\phi/2)-\tan(\phi_0/2))\b{e}_2 + M_a\b{e}_3=\b{0},\\
%\b{M}_0 -\b{M}_0^f + B(\kappa-\kappa_0)\b{e}_2=\b{0},\\
\b{U}_0^f = \b{U}_0+a \Oo\times \b{e}_3,\label{fullsys4}
\end{gather}
where $\b{U}_0$ and $\Oo$ are related to $\b{F}_0$ and $\b{M}_0$ via Eqs.~\eqref{eqU0}-\eqref{eqO0}, $\b{U}^f_0$ and $\Oo^f$ are related to $\b{F}^f_0$ and $\b{M}_0^f$ via \eqref{mobilityF}-\eqref{mobilityL}. These equations represent, in order: total force balance, total torque balance about the cell body centroid, torque balance on the hook about the basal connection point \tcb{(with the hook length assumed small)}, and the kinematic constraint of flagellum attachment to the cell body. The unknowns depend on the problem of interest.

%The force on the cell body imposed by the hook, $\b{F}_0$, and two of the components of the moment on the cell body, $\b{M}_0$, act as Lagrange multipliers to enforce the kinematic constraints of constant connectivity between the cell body and the hook, and the connection of the hook to the flagellum. Specifically, we must have that the rigid body motion of the flagellum in the $\D$ basis has $\b{U}_0^f = \b{0}$. The normal component of the moment $\b{M}_0$, however, is prescribed (the motor torque, assumed fixed and given), and is denoted by $M_a$, with $\b{M}_0\cdot \b{e}_3=M_a$. To summarize, the system of equations is

Consider first the case that the cell body position and orientation are pinned in space in the reference frame, $\U_0=\b{0}$ and $\O_0=\b{0}$ (and hence $\b{U}_0^f=\b{0}$), by an external body force $\b{F}_{ext}=\b{F}_0$ and torque $\b{M}_{ext}=\b{M}_0$ which now act as Lagrange multipliers. Equations \eqref{fullsys1}-\eqref{fullsys4} then reduce to three vector equations which are solved for the variables $\{\b{F}_0, \b{M}_0, \Oo^f\}$, resulting in the following angular velocity of the flagellum:
\begin{gather}
\O_0^f = \frac{c M_a}{8\pi\mu L^3}\left[\frac{L^2}{b^2}\cos(\phi)\b{P}-\eta \sin(\phi) \b{P}^\perp\right]-\frac{c B \eta}{4\pi \mu L^3 \ell_h}\left(\tan \left(\phi/2\right)-\tan \left(\phi_0/2\right)\right)\b{P}^{\perp\perp},
\end{gather}
with $\b{P}^\perp=\cos(\phi)\b{e}_1-\sin(\phi)\b{e}_3$, $\b{P}^{\perp\perp} = \b{e}_2$, and 
\begin{gather}\label{eq:eta}
\eta =\gamma_2-\frac{\beta_2 ^2+\beta_3 ^2}{\alpha_2}=3+\frac{3 b^2k^2}{4}\left(5-\frac{3}{k^2L^2}\right)+O(k^4b^4)\cdot
\end{gather}
The rotational velocity may instead be decomposed into its bending, precession and spin (with errors on the scale of $O(b^4k^4)$) as
\begin{gather}
\dot{\phi}=-\frac{c B \eta}{4\pi \mu L^3 \ell_h}\left(\tan \left(\frac{\phi}{2}\right)-\tan \left(\frac{\phi_0}{2}\right)\right)\label{fixedOfref2},\,\,\,\,
\dot{\theta} = \frac{c \eta M_a}{8 \pi  \mu  L^3},\,\,\,\, 
\omega =\frac{1}{8 \pi \mu  L b^2}-\frac{\eta}{8 \pi \mu L^3}.
\end{gather}
The precession rate is perhaps surprisingly independent of the bending angle, but the same viscous moment resists the lateral flagellum motion regardless of bending orientation. The force and moment required to maintain a motionless cell body do, however, depend on the bending angle.

With the body fixed in space the bending component is decoupled from the rest of the rotation, evolving until it reaches a fixed point, in this case the preferred hook angle $\phi_0$. This equilibrium hook angle is stable, per observation of \eqref{fixedOfref2}. Upon equilibration, with $\phi=\phi_0$ the external force and torque required to pin the cell body are, with $\b{P}^\perp=\cos(\phi)\b{e}_1-\sin(\phi)\b{e}_3$ and $\b{P}^{\perp\perp}=\b{e}_2$, 
\begin{gather}
\b{F}_{ext} = \frac{M_a }{L} \left\{\frac{-\beta_1}{\alpha_1}\cos(\phi_0)\b{P}+\frac{\beta_2}{\alpha_2}  \sin \left(\phi_0\right)\b{P}^\perp-\frac{\beta_3}{\alpha_2}\sin \left(\phi_0\right)\b{P}^{\perp\perp}\right\},
\end{gather}
and
\begin{multline}
\b{M}_{ext} = M_a \left\{\left(\cos(\phi_0)+\frac{a \beta_3}{L \alpha_2}\sin^2(\phi_0)\right)\b{P} -\sin\left(\phi_0\right) \left(1 -\frac{a \beta_3}{L \alpha_2}  \cos (\phi_0 )\right)\b{P}^\perp\right.\\
\left.- \frac{a}{2L \alpha_1\alpha_2}\sin\left(2\phi_0\right)\left(\alpha_2  \beta_1-\alpha_1 \beta_2\right)\b{P}^{\perp\perp}\right\}.
\end{multline}
One might expect the force required to increase with the length of the flagellum, but that assumes fixed flagellar rotation rate; with the motor torque fixed the rotation rate diminishes as $1/L$, and the resulting force is thus independent of $L$ (to leading order in the helical amplitude). Upon phase averaging over a full precession cycle, we find
\begin{gather}
\langle\b{F}_{ext} \rangle= \frac{k M_a}{2}\left(\cos^2(\phi_0)-\frac{3b^2}{2L^2}\sin^2(\phi_0)\right)\b{e}_3+O(b^4k^4),
\end{gather}
and $\langle\b{M}_{ext} \rangle= M_a \b{e}_3$, with the $b\to 0$ limit again remaining informative. With positive motor torque $M_a$ the left-handed helix is viewed as passing waves up away from the body, pushing the cell body downward and requiring an external force in the positive $\b{e}_3$ direction to maintain the cell's position. Note that Vogel \& Stark in Ref.~\cite{vs12} show the onset of a flagellum bending instability in the pinned case beyond a critical motor torque, a self-driven Euler-buckling similarities to that found in sedimenting \cite{lmss13} and axially-driven rods \cite{mk19} (see Ref.~\cite{dlns19} for a general overview). 

 \subsection{The cell body moves}\label{sec:moves}
 
Finally, we consider the full picture in which the cell body motion and flagellum motion are coupled. This amounts to the solution of the four vector equations \eqref{fullsys1}-\eqref{fullsys4} for the variables $\{\b{U}_0, \O_0, \b{U}^f_0,\Oo^f\}$. The simplest case is the axisymmetric configuration in which the preferred flagellar orientation is normal to the cell body, or $\phi_0=0$, and the bending angle is zero, $\phi=0$. Then $\dot{\phi}=0$, and extracting the precession and spin rates from Eqs.~\eqref{dphi}-\eqref{do} we obtain the translation and rotation rates
\begin{gather}
\b{U}_0 = \frac{\beta_1  M_a}{2\pi\mu L(3 a \alpha_1 X^A  +4 L c^{-1})}\b{e}_3=\frac{-c k M_a}{4 \pi  \mu  \left(3 a c X^A+2 L\right)}\left(1-\frac{b^2 k^2 L}{3 a c X^A+2 L}\right)\b{e}_3+O(b^2 k^2),\\
\Oo = -\frac{M_a}{8 \pi  a^3 \mu X^C}\b{e}_3,
\end{gather}
which are constant in time, as the body swims along in a straight path. Generally, however, for preferred bending angle $\phi_0$ and instantaneous bending angle $\phi$, the body translation and rotation rates depend on the complete body and flagellum geometries (see Appendix A). But using $\dot{\phi} = \b{e}_2 \cdot \left(\Oo^f-\Oo\right)$ they reveal an equation describing the dynamics of the bending angle alone (even while the entire system translates and rotates and the flagellum precesses):
\begin{gather}\label{eq: phidotmain}
%\dot{\phi} = -\frac{B}{24\pi \mu  L^3 \ell_h(3\alpha_2+\lambda c^{-1} Q_3 )\nu} \left(C_1(\phi)\left[\tan\left(\frac{\phi}{2}\right)-\tan\left(\frac{\phi_0}{2}\right)\right]-\tilde{M}_a \lambda \sin(\phi)C_2(\phi)\right),
\dot{\phi} = -\frac{B}{8\pi \mu  L^3 \ell_h} \left(C_1(\phi)\left[\tan\left(\frac{\phi}{2}\right)-\tan\left(\frac{\phi_0}{2}\right)\right]-\tilde{M}_a  \sin(\phi)C_2(\phi)\right),
\end{gather}
where $\tilde{M}_a = (B/\ell_h)^{-1}M_a$ is a dimensionless active moment and $\lambda=L/a$ is the scaled flagellum length, and the functions $C_1(\phi)$ and $C_2(\phi)$ are given in Appendix B. In the slender flagellum limit \tcb{($c\to \infty$)}, however, we have
\begin{gather}\label{c1c2old}
C_1(\phi) =2c \eta +\frac{1}{Y^C}\left[2\lambda ^3+5\lambda^2  m_3\cos(\phi)+3m_3^2\lambda Y^C\left(Q_2+Q_1\cos(2\phi)\right)\right]+O(\lambda/c),\\
C_2(\phi)=\frac{\lambda^2 m_2}{Y^C}\cos(\phi)+\lambda Q_1 m_3 \left(m_1+m_2\cos(2\phi)\right)+O(\lambda/c),
\end{gather}
where $\eta$ is given in Eq.~\eqref{eq:eta} and we have defined
\begin{gather}
m_1 = -\left(\frac{\beta_1}{\alpha_1}+\frac{\beta_2}{\alpha_2}\right) = \frac{k L}{2}\left(1-\frac{3 b ^2}{2 L^2}\right)+O(b^4k^4),\\
m_2 = -\left(\frac{\beta_1}{\alpha_1}-\frac{\beta_2}{\alpha_2}\right) =\frac{k L}{2}\left(1+\frac{3 b ^2}{2 L^2}\right)+O(b^4k^4) ,\\
m_3 = -\frac{2\beta_3}{3\alpha_2} =1+\frac{1}{2}b ^2k^2 \left(1-\frac{3}{2 (kL)^2}\right) +O(b^4k^4),\\
Q_1 =\frac{3}{4Y^C}+\frac{1}{Y^A}-\frac{1}{X^A},\,\,\,\,\, Q_2 = \frac{1}{X^A }+\left(1+2\left(\frac{\beta_2}{\beta_3}\right)^2\right) \left(\frac{1}{Y^A }+\frac{3}{4Y^C }\right).\label{eq:Q1Q2}
\end{gather}
In the limit of a very large cell body or very short flagellum, $\lambda\to 0$, the problem reduces to that studied in the previous section: the base of the flagellum is effectively pinned on a wall, $C_1(\phi;\phi_0)=2c \eta$ and $ C_2(\phi)=0$, and $\phi$ rapidly relaxes to its preferred value of $\phi_0$. \tcb{That an equation for the bending angle alone must emerge is clearest in the reference frame - since the body is axisymmetric the body may spin in the reference frame about the $\b{e}_3$ direction, but the bending angle rate $\dot{\phi}$ is wholly independent of the azimuthal angle $\theta$, and no other physics in the problem select any special orientations or directions.} 

Substantial simplification ensues when considering nearly spherical bodies and small flagellar helical amplitudes, leaving simply
\begin{gather}
C_1(\phi)=\frac{1}{\nu(\phi)}\left(6 c+\frac{2 \lambda ^2}{3c} \left(3 \lambda ^2+6 \lambda  \cos (\phi )+14\right)+\frac{\lambda}{2}\left(4 \lambda ^2+12 \lambda  \cos (\phi )+3 (11+\cos (2 \phi))\right)+\frac{4 \lambda ^5}{9 c^2}\right)+O(b^2k^2,e^2),\\
C_2(\phi)=\frac{k L \lambda \cos (\phi)}{12\nu(\phi)}\left(9\cos(\phi)+6\lambda+2c^{-1}\lambda^2\right)+O(b^2k^2,e^2),
\end{gather}
with $\nu(\phi)=1+c^{-1}\lambda(11-\cos(2\phi))/8+7c^{-2}\lambda^2/18+O(b^2k^2,e^2)$, as $bk\to0$ and $e\to 0$.

%\begin{gather}
%C_1(\phi) = \left(6c+\lambda  \left(\frac{6}{X^A }+\frac{3 Q_3 }{2}+2 Q_1  \cos (2 \phi )\right)+\frac{2 \lambda ^3}{Y^C }+\frac{6 \lambda ^2 \cos (\phi )}{Y^C }\right)+O(b^2k^2,c^{-1}),\\
%C_2(\phi) = \frac{(k L \cos (\phi ))}{6} \left(\frac{\lambda ^2}{c (X^A  Y^C )}+3 \left(2 Q_1  \cos (\phi )+\frac{\lambda }{Y^C }\right)\right)+O(b^2k^2,c^{-1}),
%\end{gather}
%and finally for nearly spherical bodies,

%Representations of these coefficients in terms of the flagellum geometry are included in Appendix A. 

\subsubsection{Straight swimming is unstable beyond a critical active moment}

\begin{figure}[htbp]
\begin{center}
\includegraphics[width=.9\textwidth]{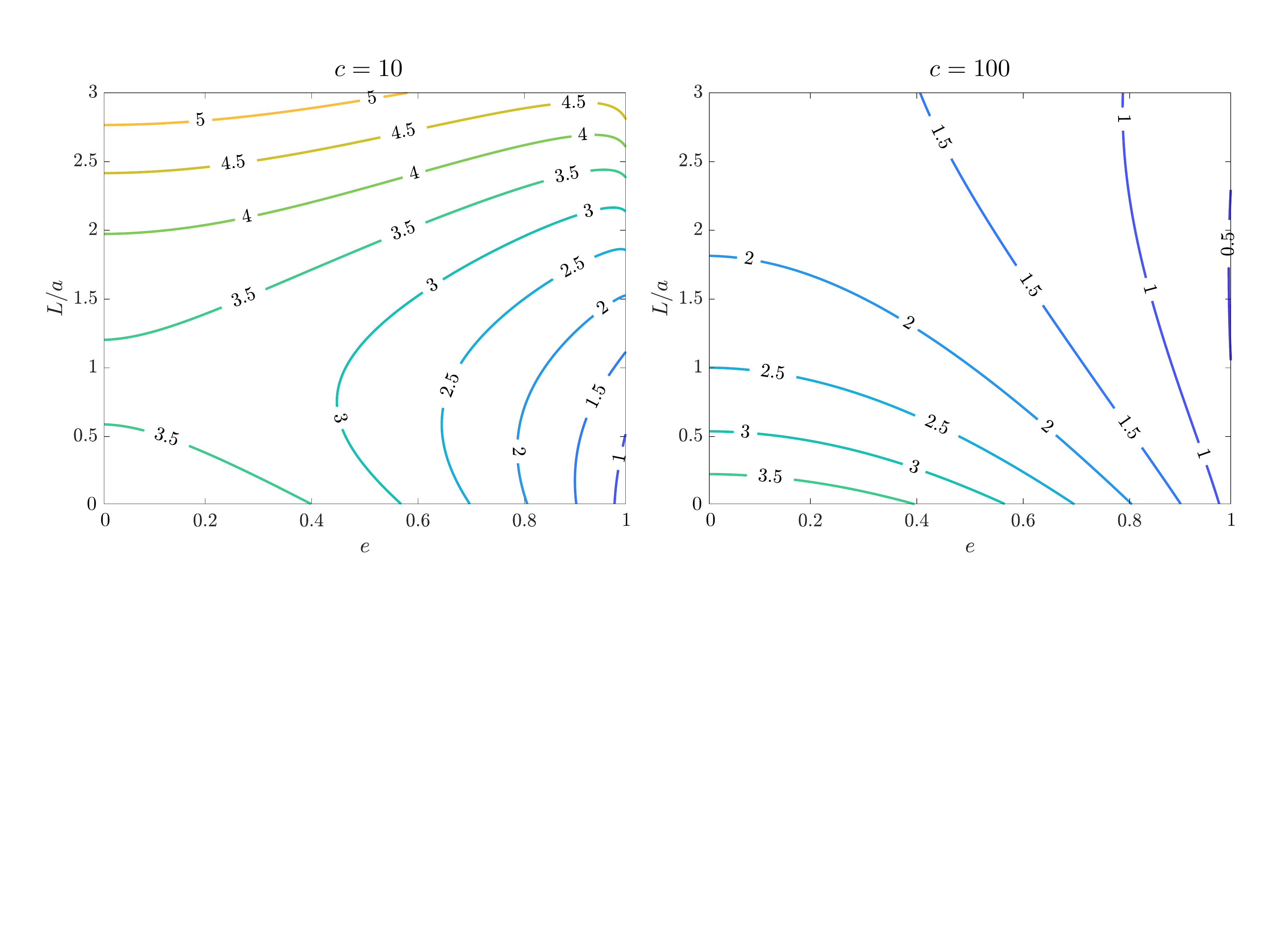}
\caption{Contours of the dimensionless function $\mathcal{F}$ are shown, where $e$ is the cell body eccentricity and $M_a^*=[(B a c)/(k L^2 \ell_h)]\mathcal{F}$ is the critical active moment beyond which a straight swimming trajectory is unstable (to leading order in the flagellar amplitude). The critical active moment is generically non-monotonic in $L/a$.}
\label{fig: M_crit}
\end{center}
\end{figure}

Is the straight swimming configuration, with $\phi_0=0$ and $\phi=0$, stable? Expanding Eq.~\eqref{eq: phidotmain} about small $\phi$ (without assuming $\lambda/c$ to be small), we find the linearized equation
\begin{gather}
\dot{\phi} = -\frac{B}{8\pi \mu L^3 \ell_h}\left(\frac{1}{2}\tilde{C}_1- \tilde{M}_a \tilde{C}_2\right)\phi,
\end{gather}
with
\begin{gather}
\tilde{C}_1 = \frac{1}{3\alpha_2+\lambda c^{-1}Q_3 }\left(6 c \alpha_2 \eta +2 \lambda \gamma_2 Q_3  +\frac{6}{Y^C}\left(-2 \beta_3\lambda^2+\alpha_2 \lambda^3\right)+\frac{8 \lambda^4}{c Y^A Y^C}\right),\\
\tilde{C}_2 = \frac{\lambda}{(\alpha_1X^A +4\lambda (3c)^{-1})(3\alpha_2+\lambda c^{-1}Q_3 )}\left((X^A Q_3 -4)\beta_1\beta_3+\frac{3\lambda X^A}{Y^C}\alpha_1 \alpha_2 m_2+\frac{4\lambda^2(\beta_2-\beta_1)}{c Y^C}\right),
\end{gather}
having defined $Q_3  = 3/Y^C+4/Y^A$. A critical orientational instability is thus observed when the active dimensionless motor torque $\tilde{M}_a=(B/\ell_h)^{-1}M_a$ crosses a threshold $\tilde{M}_a^*:= \tilde{C}_1/(2 \tilde{C}_2)$, or
\begin{gather}
\tilde{M}_a^*=\frac{(\alpha_1X^A +4\lambda (3c)^{-1})\left(6 c \alpha_2 \eta +2 \lambda \gamma_2 Q_3  +(2/Y^C)\left(-6 \beta_3\lambda^2+3\alpha_2 \lambda^3+4 c^{-1}\lambda^4/Y^A\right)\right)}{(X^A Q_3 -4)\beta_1\beta_3+3\lambda X^A\alpha_1 \alpha_2 m_2/Y^C+4c^{-1}\lambda^2(\beta_2-\beta_1)/Y^C}.
\end{gather}
In the small helical amplitude limit $bk \to 0$, this may be written as
\begin{gather}\label{Mastar}
\tilde{M}_a^*=\frac{c}{a k \lambda^2}\mathcal{F}(\lambda,e,c)+O(b^2k^2)
\end{gather}
where
\begin{gather}\label{eq:F}
\mathcal{F}(\lambda,e,c) =\frac{4 \left(3 X^A+2 \lambda c^{-1}\right) \left(9 Y^A  Y^C+3 \lambda c^{-1} \left(\left(\lambda ^2+3 \lambda +3\right) Y^A +4 Y^C \right)+\lambda ^4c^{-2} \right)}{9 [(3+2 \lambda) X^A  Y^A +4 X^A  Y^C -4 Y^A  Y^C]+6 \lambda ^2 Y^A c^{-1}}.
\end{gather}
Figure~\ref{fig: M_crit} shows the contours of $\mathcal{F}$ for $c=10$ and $c=100$ across a range of flagellum lengths, $\lambda=L/a$, and body eccentricities, $e$. When the flagellum is substantially aspherical ($e$ increasing towards $1$), straight swimming tends to be less stable in the sense that the critical active moment before instability is reduced. As shown in the leftmost panel with $c=10$, the critical active motor torque is generically non-monotonic in the flagellum length, as described by Nguyen \& Graham \cite{ng17}. As $\lambda \to 0$, $\tilde{M}_a^*\sim (3c)/(ak Q_1 \lambda ^2)$ as $\lambda\to 0$, a singularity indicating that the active motor torque required to destabilize the system grows without bound as the relative flagellum length goes to zero (or the cell body size becomes large). This is consistent with the limiting case studied in the previous section. Meanwhile, for large $\lambda$, $\tilde{M}^* \sim 4 \lambda/(3 a c k Y^A)$ as $\lambda \to \infty$, and once again in this limit the motor torque required to destabilize the system becomes very large. Note that the non-monotonicity of the critical active moment $\tilde{M}_a^*$ is not observed if terms of order $c/\lambda$ are not retained. For yet a clearer inspection, for a spherical cell body we find
\begin{gather}\label{eq: critM}
\tilde{M}_a^*=\left(\frac{4 (3c+2 \lambda) }{3a k\lambda^2}\right)\frac{9+3  c^{-1} (\lambda^3+3\lambda^2+7\lambda) +\lambda ^4 c^{-2}}{3(3+2 \lambda)+2 \lambda ^2 c^{-1}},
%\tilde{M}_a^* &=\frac{4 c \eta +\lambda  \left(\frac{7}{3} (m_1 -m_2 )^2+4 \lambda ^2+10  m_3 \lambda+21 m_3 ^2 \right)}{3 \lambda  m_3  (m_1 +m_2 )+4 m_2  \lambda ^2}+O(c^{-1}) \,\,\,\,\mbox{as} \,\,\,\, c \to \infty\\
%&=\frac{1}{kL}\left(\frac{12 c+4 \lambda ^3+10 \lambda ^2+21 \lambda }{2 \lambda ^2+3 \lambda }\right)+O(c^{-1},b^2k^2) \,\,\,\,\mbox{as} \,\,\,\, c \to \infty,\,\, bk\to 0.
\end{gather}
from which the non-monotonicity in $\lambda$ is more easily argued.

For a very slender flagellum $c\gg 1$ (e.g. with $c=100$ in the right-most panel of Fig.~\ref{fig: M_crit}), a longer flagellum results in greater orientational instability since the moment applied at its base by the flagellar hook is fixed, but the viscous resistance to realigning the flagellum to the normal direction is increased. However, the stability of this configuration also depends on the flagellum slenderness. As $c \to \infty$ the normal orientation is stable, as the resistance to returning the filament to the normal direction is reduced. In practice, however, $c$ is logarithmic in the filament slenderness, and $c$ is more likely to be on the scale of $10-100$ \cite{bw77}. At leading order for large $c$ we do not observe a dependence of stability on the flagellum amplitude, but we do observe a dependence on the flagellum wavenumber, owing to the discussion at the end of \S IV(c). The hook is stabilized with a smaller wavenumber $k$ or a longer flagellar wavelength (via the prefactor $c/(a k \lambda^2)$ in $\tilde{M}_a^*$), since a given motor torque generates less thrust in that case, all else being equal. Chirality is important, however: changing the sign of the active moment results in a reversal in the flagellum rotation direction, rendering the unstable regimes above stable; namely, $\phi=0$ is always a stable equilibrium in this model if the flagellum is pulling on the cell body. For a slender cell body $e \approx 1$ we find $\tilde{M}_a^* \sim 4 c a/[k L^2 ((L_e-1)\lambda+L_e)]+O(1)$ as $e \to 1$, further highlighting the competition between the slenderness of the flagellum and the slenderness of the cell body.

%\tcb{Remaining in the slender flagellum regime, the critical active moment required to destabilize the normal orientation in Eq.~\eqref{eq: critM} decreases as a function of $\lambda=L/a$ for biologically relevant parameters. This trend lies in apparent contrast to the buckling instability identified by Vogel \& Stark, who describe a critical motor torque before flagellum buckling which scales instead as $1/a$ for large $a$ \cite{vs12}. The result above focuses attention on the relevance of hook flexibility, rather than flagellum flexibility; which effect dominates will be system dependent. In the limit of a very large cell body, $\lambda\to 0$, the normal orientation of the flagellum is stable for all active moments, as the problem reduces to that studied in the previous section wherein the base of the flagellum is pinned to an effective wall. For a slender cell body $e \approx 1$ we find $\tilde{M}_a^* \sim 4 c a/[k L^2 ((L_e-1)\lambda+L_e)]+O(1)$ as $e \to 1$, further highlighting the competition between the slenderness of the flagellum and the slenderness of the cell body.}

To summarize, if the preferred flagellum orientation is normal to the cell surface, a straight swimming path is predicted for motor torques below a critical value, including negative values corresponding to puller-type swimming. For motor torques beyond a critical value, which is generically non-monotonic in the flagellum length, the normal orientation is unstable, and the hook angle $\phi$ must equilibrate at another value \tcb{(perhaps in contact with the cell body)}, as dictated by the autonomy of the dynamical system. In that configuration, the trajectory is assured to be helical, given the discussion in \S II, a point to which we will shortly return. 

\subsubsection{Arbitrary preferred hook angle and helical trajectories}
\begin{figure}[htbp]
\begin{center}
\includegraphics[width=.95\textwidth]{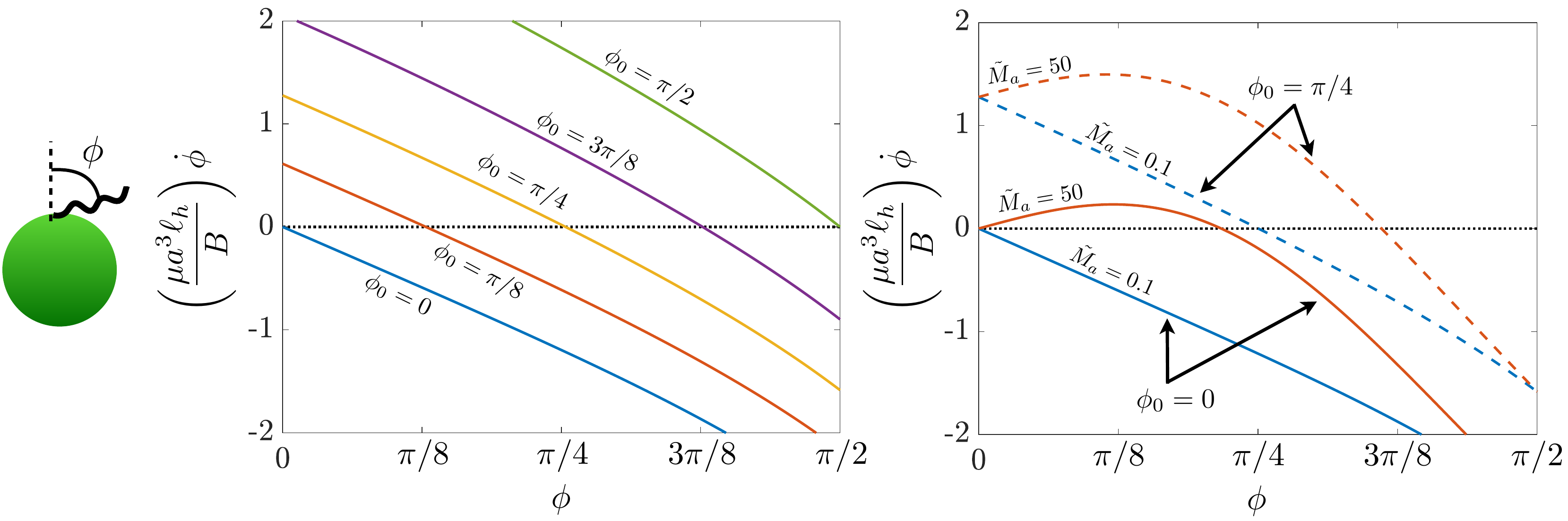}
\caption{$e=0$. (Left) The dimensionless hook bending rate as a function of the current hook angle is shown for a selection of preferred hook angles, $\phi_0$, with dimensionless active motor torque, $\tilde{M}_a=1$. The body is spherical, $e=0$, and the remaining dimensionless parameters are $c=10$, $k L=1$, and $\lambda=L/a=1$. (Right) The same, but for two different values of the dimensionless active motor torque $\tilde{M}_a$, and two different preferred hook angles.}
\label{fig: phidots}
\end{center}
\end{figure}
%\begin{multline}\label{dotphic}
%\dot{\phi}=\frac{3B}{4 \pi \mu L^3 \ell_h}\left\{\frac{\tilde{M}_a  \lambda k L}{4}\left(n_1\cos(\phi)+\lambda n_2\right)\sin (2 \phi)\right.\\
%\left.-\left(c+2 \lambda ^3 n_2+5 \lambda ^2 n_2 \cos (\phi)+\lambda  \left(\frac{3}{2} n_1\cos (2 \phi)+n_4\right)\right)\left(\tan \left(\frac{\phi}{2}\right)-\tan \left(\frac{\phi_0}{2}\right)\right)\right\},
%\end{multline}
%where
%\begin{gather}
%n_4=\frac{\left(15+36 e^2-15 e^4\right) L_e-6 e\left(5-e^2\right)}{32 e^3 \left(2-e^2\right)}
%\end{gather}
%is an increasing function of the body eccentricity $e$ with $n_4=11/8$ as $e\to 0$. 
Finally we consider the case that the preferred flagellar orientation is not normal to the cell, $\phi_0\neq 0$. Figure~\ref{fig: phidots} (left) shows the dimensionless hook bending rate $\dot{\phi}$ as a function of the current bending angle $\phi$ for a selection of preferred angles $\phi_0$, with dimensionless active motor torque, $\tilde{M}_a=(B/\ell_h)^{-1}M_a=1$. For these plots the body is assumed spherical, $e=0$, and the remaining dimensionless parameters are chosen as follows: $c=10$, $k L=1$, and $\lambda=1$. An angle $\phi^*$ close to (but not generally equal to) $\phi_0$ is observed to be stable. But upon increasing the motor torque, $\dot{\phi}$ can become non-monotonic, as shown in Fig.~\ref{fig: phidots} (right), which includes two preferred angles, $\phi_0=0$ and $\phi_0=\pi/4$, and two dimensionless motor torques $\tilde{M}_a=0.1$ and $\tilde{M}_a =50$. The straight swimming case, studied in the previous section, reveals a pitchfork bifurcation beyond a critical $\tilde{M}_a$ in between these values (or a supercritical Hopf bifurcation if viewed as a two-dimensional system describing the flagellum tip position). \tcb{A new stable fixed point emerges in a continuous departure from zero just as soon as the active moment increases beyond the critical value $\tilde{M}_a^*$, and $\phi=0$ becomes an unstable fixed point.} If $\phi_0>0$, however, the non-monotonicity of $\dot{\phi}$ does not affect the stability of the fixed point, it merely nudges the fixed point closer to $\pi/2$. 
\begin{figure}[htpb]
\center
	\includegraphics[width=0.6\textwidth]{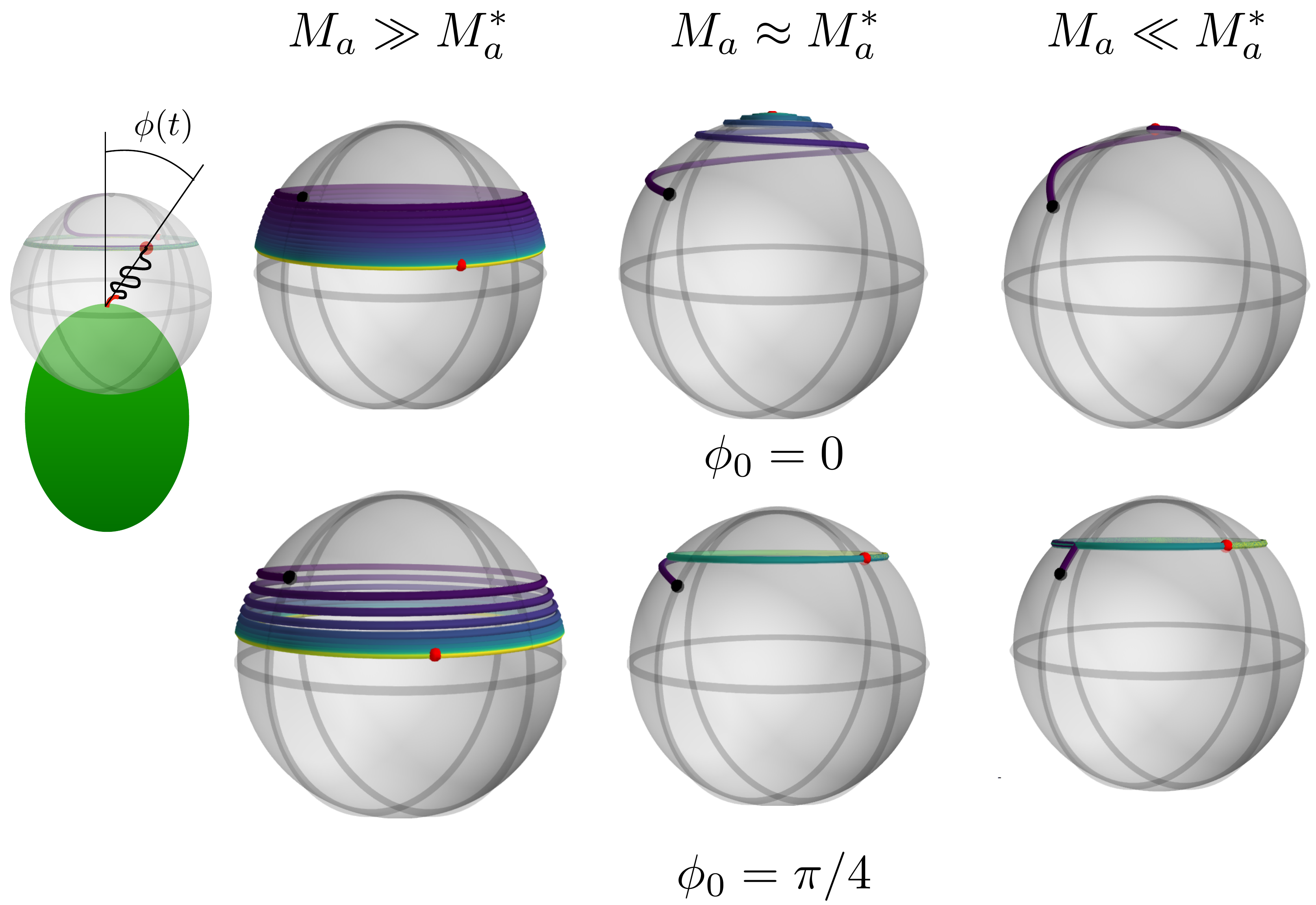}
	\caption{The dynamics of the flagellar tip, represented as motion on a sphere (see schematic on the left). The preferred hook angle is $\phi_0=0$ (top row) or $\phi_0=\pi/4$ (bottom row). Left column: large motor torque applied to the flagellum results in flagellum orientation instability and the tip slowly winds down towards a large equilibrium hook angle. Center column: with the motor torque very near to but below the critical motor torque for instability, the flagellum orientation relaxes to an equilibrium after a few precessions. Right column: for very small motor torques, the bending stiffness of the hook overwhelms any other effects and the orientation returns to the preferred angle exponentially fast in time. The dynamics of the tip are consistent with a supercritical Hopf bifurcation. }
	\label{FlagellumPaths}
\end{figure}

Another view of the dynamics is provided by focusing on the path taken by the tip of the flagellum in a frame in which the body is fixed. The top row of Fig.~\ref{FlagellumPaths} shows the dynamics of the flagellar tip when $\phi_0=0$ for an active moment substantially larger than the critical moment, one commensurate with but just below that critical value, and one which is much smaller than the critical value. The unstable case involves many rotations and a slow departure away from the normal direction, the commensurate case shows oscillations with exponential decay towards the equilibrium, and the final case reveals an overdamped decay to the preferred orientation. The second row of Figure~\ref{FlagellumPaths} shows the flagellum tip dynamics with a preferred hook angle of $\phi_0=\pi/4$, showing similar dynamical structure as in the $\phi_0=0$ case. When the active moment is small the hook angle rapidly adjusts to nearly the preferred angle, $\phi_0$, then the flagellum precesses around the normal direction with a constant rotation rate. 
\begin{figure}[htbp]
\begin{center}
\includegraphics[width=\textwidth]{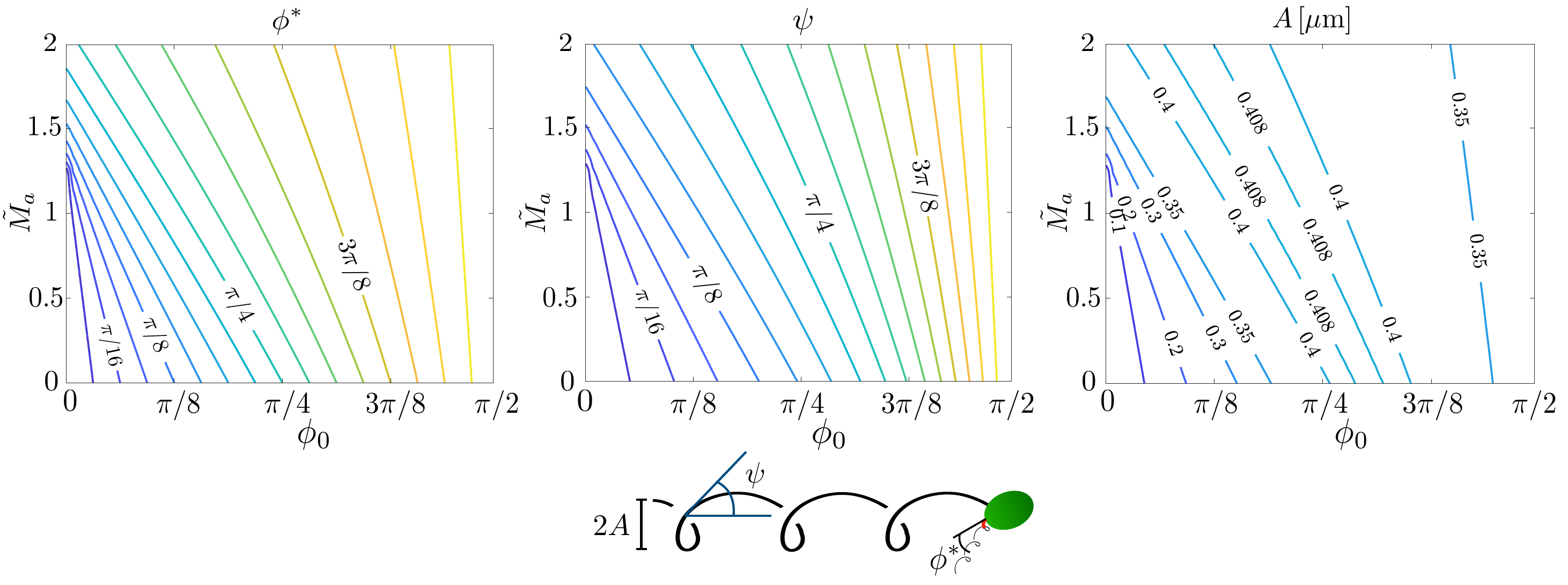}
\caption{Helical swimming trajectories using physical constants measured for {\it P. aerugenosa}; $b=0.2\mu$m, $a=2\mu$m, $\e\approx 1/200$ ($c=10$), $k=3.34\mu$m$^{-1}$, $L=4\mu$m, $b/a=1/4$. (Left) Equilibrium hook angle; (center) helical trajectory pitch angle, $\psi$; (right) helical trajectory amplitude, $A$. The phase space of spontaneous hook angle and motor torque show distinct regions of hook angle instability, but also an intermediate range of values of the hook angle for which large motor torques might still be used without destabilizing the flagellum orientation.}
\label{fig: Trajectories}
\end{center}
\end{figure}

%\begin{gather}
%\mathcal{J}(\phi_0,L/A;e\to 1) = 2 \sin \left(\frac{\phi_0 }{2}\right) \cos ^3\left(\frac{\phi_0 }{2}\right) \left(\log \left(\frac{2}{1-e}\right) (L/A +\cos (\phi_0 ))-L/A \right)+O(1-e),\\
%\mathcal{J}_2(e\to 1) = -\frac{\sin (\phi_0 ) \cos ^2\left(\frac{\phi_0 }{2}\right) \left(9 a \left(k^2 L^2-1\right) \cos (\phi_0 )+9 a+10 k^2 L^3-6 L\right)}{12 a}+O(1-e).
%\end{gather}

Once the dynamical system above has selected the fixed point, $\phi^*$, such that $\dot{\phi}=0$, the body moves along a helical trajectory. Performing a simple numerical root-finding algorithm on Eq.~\eqref{eq: phidotmain} and evaluating $\Uo$ and $\Oo$ at this angle, the helical amplitude $A$, and the pitch angle $\psi$ are quickly computed using the expressions in \S\ref{sec:trajec}. Figure~\ref{fig: Trajectories} shows contours of the equilibrium bending angle, amplitude, and pitch angle as functions of the dimensionless motor torque $\tilde{M}_a=(B/\ell_h)^{-1}M_a$ and the preferred hook angle $\phi_0$. To generate these plots we use biologically relevant material parameters, taking values describing an organism with a single polar flagellum, {\it P. aeruginosa}, from Refs.~\cite{fsa08,vwmlksor14}, and assuming the `normal form' of the flagellum. The cell body has half-length $a=1\mu$m and aspect ratio $4$, giving an eccentricity $e=\sqrt{15}/4\approx 0.97$. The relative flagellum length is taken to be $\lambda=L/a=2$, the helical radius is $b = 0.2 \mu$m and the pitch is $P=1.4 \mu$m, so we take $k= 2\pi/\sqrt{P^2+4\pi^2b^2}=3.34\mu$m$^{-1}$. The flagellum aspect ratio is $\e= 20 n$m$/4\mu$m $=0.005$, which gives $c=-\log(\e^2)-1\approx 10$.

The equilibrium pitch angle is an increasing function of the motor torque and is only non-smooth (but still continuous, consistent with Ref.~\cite{ng17} for $\phi_0=0$. Large motor torques nudge the equilibrium angle towards $\psi=\pi/2$, and the flagellum closer to the tangent plane of the cell body at the hook connection point. The helical pitch angle transitions from small (nearly straight swimming) for $\phi_0=0$ and small motor torque to values closer to $\psi=\pi/2$ (and a tightly coiled path) for large preferred hook angles. The pitch angle of the helical trajectory tends to be smaller than the equilibrium hook bending angle, as it incorporates the full hydrodynamic coupling with the cell body. Finally, the helical amplitude shows a remarkable feature. As soon as the equilibrium hook angle is not very small, either because the preferred angle is not small, or because the motor torque is large, the helical amplitude remains in a very small range around a maximum near $A\approx 0.6\mu$m. We might expect, then, to find {\it P. aeruginosa} cells swimming along a variety of helical paths with different pitch angles but very similar amplitudes. 

\section{Discussion}
We have aimed to isolate the role of a flexible flagellar hook in microorganism locomotion by studying a model phase-averaged rigid helical flagellum. The mathematical simplicity offered by this modeling choice allowed for the analytical prediction of a critical stability criterion for the motor torque beyond which the flagellum hook angle becomes unstable, a prediction which included details about the flagellum geometry. This constraint also indicates that the straight- or nearly-straight swimming of some flagellated microorganisms might not be possible beyond a maximum speed, as increasing the motor torque to increase the thrust may destabilize the trajectory. This is reminiscent of vastly larger organisms like fish, which exploit flexible appendages for speed and efficiency benefits, but beyond a critical flapping frequency the appendage deformability can result in dramatically poor performance, even retrograde motion \cite{ln96,aml07,smsz10}. The constraints on the flagellar hook flexibility, and the functionality conferred by its deformability, may lead to deeper insights about the evolutionary development of flagellar cell motility. Observations of natural motility continues to motivate the creation of synthetic analogues \cite{Ebbens2010,Nelson2010,Sengupta2012,Li2017,Ricotti2017,Hu2018,Tsang2020}, bio-hybrids \cite{Carlsen2014}, and even energy harvesting applications \cite{zs20}. Other important questions relate to enhanced diffusion, fluid transport and mixing \cite{kb04,lggpg09,ltc11,Thiffeault15}; the role of hook flexibility and flagellar buckling on such phenomena is not currently clear. Many organisms swim through fluids which are substantially non-Newtonian, and the additional physical forces conferred on the helical flagellum (and the cell body) by viscoelasticity \cite{fpw07,fwp09,ls15,lpb11,slp13,msrwmp14,zy19,bphs20} and shear-dependent viscosity \cite{gglz17,dldp20}, for instance, are likely to contribute to these buckling criteria as well, just as they can affect the dynamics of flagellar bundling \cite{qb20}.

\tcb{We have only considered a phase-averaged model flagellum in this work, but it is worth considering what possible effects a true finite-length flagellum might introduce (while still neglecting the possibility of hard contact between the flagellum and cell body; see Refs.~\cite{ng18,pkl19} for such simulations). Each of the mobility/resistance coefficients in Sec.\ref{sec: model flagellum} generally depends upon the flagellar phase; hence instead of arriving at a single equation for the bending angle, Eq.~\eqref{eq: phidotmain}, a coupled two-dimensional dynamical system linking the bending angle and the phase, $(\phi,\delta)$, would emerge (ostensibly a rather ungainly one). But the system would remain autonomous, reducing the space of possible dynamical behaviors via the Poincar\'e-Bendixson theorem, and convergence to a limit cycle (a periodic orbit) is thus still expected. If the cell body is not axisymmetric, however, the flagellum state is governed by a three-dimensional dynamical system with phase space $(\phi,\delta,\theta)$, opening up a much wider space of possibilities for the flagellum tip dynamics. The net effect this may have on the ultimate swimming trajectory is hard to predict, but such body shapes can enhance cell motility \cite{lgmtpb14}.}

Among our findings we have shown that straight-swimming is less stable for longer flagella or for cell bodies with a smaller aspect ratio, in the sense that the active motor torque required to destabilize straight swimming is decreased. On the other hand, when the flagellum is vanishingly thin, there is very little resistance to returning the flagellum to the preferred hook angle. The final outcome depends on all parameters, including the relative flagellum length and helical wavenumber, and the cell body eccentricity. But once the hook angle settles to its equilibrium, the cell body still traverses a helical path, even as the flagellum precesses and the cell body translates and rotates. The stability of the body trajectory is substantially softened when the preferred hook angle is non-zero, in the sense that the resulting helical trajectory appears to have a much smoother dependence on the active moment applied to the hook at the base of the flagellum. This raises the question of whether the non-zero spontaneous hook angle may have arisen in part to confer a more predictable path through the fluid, predictability upon which other biological functions might develop like flow-enhanced molecular transport \cite{ssgpkg06}, albeit along a dizzying helical trajectory. 

\appendix

\section{Derivation of Eq.~\eqref{eq:kappaderiv}}

We will use $\mathcal{F}(\b{a}_1,\b{a}_2,\b{a}_3)$ to denote the determinant of the matrix with columns $\b{a}_1, \b{a}_2, \b{a}_3$. Consider $\left(\dot\r\times \ddot\r\right) \cdot \b{c}$ for some vector $\b{c}$. Using $\D\D^T = \I$ and $\det(\D) = 1$, we have
\begin{gather}
\left(\left(\D \cdot\b{a}\right) \times \left(\D \cdot \b{b}\right)\right)\cdot \b{c}= \mathcal{F}\left(\D \cdot\b{a}, \D \cdot \b{b},\, \b{c} \right)
=\mathcal{F}\left(\D \cdot\b{a}, \D \cdot \b{b}, \D\D^T \cdot \b{c} \right)= \det\left(\D\right) \mathcal{F}\left(\b{a} , \,\b{b},\D^T \cdot\b{c} \right)\nonumber\\
=\mathcal{F}\left(\b{a},\, \b{b}, \D^T \cdot \b{c} \right) = (\b{a}\times \b{b}) \cdot (\D^T \cdot \b{c})  =\left(\D \cdot (\b{a} \times \b{b})\right)\cdot \b{c}\nonumber.
\end{gather} 
Since $\b{c}$ is an arbitrary vector, we find $\left(\D \cdot\b{a}\right) \times \left(\D \cdot \b{b}\right) = \D\cdot (\b{a} \times \b{b})$.

\section{Cell translation and rotation rates}

In the full problem of a freely swimming cell body, solving Eqs.~\eqref{fullsys1}-\eqref{fullsys4} we find the body translation in the reference frame,
\begin{multline}
\b{U}_0=\frac{-M_a}{12\pi \mu a L X^A}\left(\frac{X^A}{Y^A} m_2\sin(2\phi) \b{e}_1+3\frac{X^A}{Y^A} m_3\sin(\phi)\b{e}_2+\left(m_1+m_2\cos(2\phi)\right)\b{e}_3\right)\\
+\frac{B}{6 \pi \mu a L \ell_h X^A}\left\{3\frac{X^A}{Y^A}m_3\left(\sin (\phi )-\tan \left(\frac{\phi_0}{2}\right)\cos(\phi)-\tan \left(\frac{\phi }{2}\right)\right)\b{e}_1+(m_1-m_2)\frac{X^A}{Y^A}  \left(\tan \left(\frac{\phi }{2}\right)-\tan \left(\frac{\phi_0 }{2}\right)\right)\b{e}_2\right.\\
\left.-3m_3\left(1-\cos (\phi )-\tan \left(\frac{\phi_0 }{2}\right) \sin (\phi )\right)\b{e}_3\right\}+O(\lambda/c),
\end{multline}
and the rotation rate,
\begin{multline}
\Oo = \frac{-M_a}{8\pi \mu a^3 X^C}\left(-\frac{3m_3 X^C}{2\lambda Y^C}\sin(\phi)\b{e}_1 +\frac{m_2 X^C}{2\lambda Y^C}\sin(2\phi) \b{e}_2+\b{e}_3 \right)\\
+\frac{B}{8\pi \mu a^2 L \ell_h Y^C}\left(\tan \left(\frac{\phi}{2}\right)-\tan \left(\frac{\phi_0 }{2}\right)\right)\left((m_2-m_1)\b{e}_1+\left(2\lambda+3m_3\cos(\phi)\right)\b{e}_2\right)+O(\lambda/c),
\end{multline}
as $c/\lambda \to \infty$, where we have defined the scaled flagellum length $\lambda=L/a$ and $m_1$, $m_2$, and $m_3$ are given in \S~\ref{sec:moves}. The component of the flagellum rotation rate relevant to the bending angle dynamics is 
\begin{multline}
\b{e}_2 \cdot \Oo^f  = \frac{M_a  Q_1}{8 \pi \mu a L^2 }m_3\sin (\phi)\left(m_1+m_2\cos (2 \phi )\right)\\
-\frac{B}{8 \pi \mu \ell_h L^3 }\left(2c \eta +\frac{3\lambda^2  m_3}{Y^C}+3\lambda m_3^2(Q_2+Q_1\cos(2\phi))\right)\left(\tan \left(\frac{\phi }{2}\right)-\tan \left(\frac{\phi_0 }{2}\right)\right)+O(\lambda/c),
%\b{e}_2 \cdot \Oo^f = \frac{k M_a}{32\pi \mu a L X^A Y^A Y^C} \cos^2(\phi)\sin(\phi)\left(3 X^A Y^A+4 Y^C(X^A -Y^A)\right)\\
%-\frac{3 B \tan(\phi/2)}{32\pi \mu L^3\ell_h X^A Y^A Y^C}\left( 8 c X^A  Y^A  Y^C +4 \lambda ^2 X^A  Y^A  \cos (\phi)+3 \lambda  X^A  Y^A +\lambda  \cos (2 \phi) (3 X^A  Y^A +4 X^A  Y^C -4 Y^A  Y^C )+4 \lambda  Y^C  (X^A +Y^A )\right)
\end{multline}
with $Q_1$ and $Q_2$ defined in \eqref{eq:Q1Q2} and $\eta$ given in \eqref{eq:eta}. Although $c/\lambda=ca/L$ is assumed large for the expressions above, no such assumption is made when presenting the hook bending rate, Eq.~\eqref{eq: phidotmain}, and we find
\begin{gather}\label{c1c2}
C_1(\phi) =\frac{c M_1+M_0+ c^{-1}M_{-1}+c^{-2}M_{-2}+c^{-3}M_{-3}}{3\alpha_1\alpha_2 X^A X^C Y^C(3\alpha_2+c^{-1}\lambda Q_3 )\nu(\phi)},\\
C_2(\phi) = \frac{\lambda\left(N_0+ c^{-1}N_{-1}+c^{-2}N_{-2}\right)}{3(3\alpha_2+c^{-1}\lambda Q_3 )\nu(\phi)},
\end{gather}
where $Q_3  = 3/Y^C+4/Y^A$ and
\begin{gather}
\nu(\phi) =1+\frac{\lambda}{6c}  \left(\left(\frac{1}{\alpha_2}+\frac{1}{\alpha_1}\right)\left(Q_3  +\frac{4}{X^A }\right)+4\left(\frac{1}{\alpha_2}-\frac{1}{\alpha_1}\right)Q_1 \cos (2 \phi )\right)+\frac{4\lambda^2}{9c^2}\frac{Q_3  }{\alpha_1  \alpha_2  X^A},
\end{gather}
and the coefficients are as follows: 
\begin{align}
M_1 =&\, 18 \left(\alpha_2 \eta\nu(\phi)-\beta_3^2 (1-\nu(\phi))\right),\\
M_0 =&\, 18 \lambda ^3 \alpha_1  \alpha_2 ^2 X^A  Y^A -36  \lambda ^2 \alpha_1  \alpha_2  \beta_3 X^A  Y^A  \cos (\phi )\nonumber\\
&+3 \lambda  Y^A  Y^C\left[4 \alpha_2  \beta_3 ^2 Q_1  X^A   \cos (2 \phi )-Q_3  X^A    \left((2 \alpha_1 +\alpha_2 )\beta_3 ^2 -2 \alpha_1  \alpha_2  \gamma_2  \nu(\phi) \right)-4 \alpha_2  \beta_3 ^2 \right],\\
M_{-1} =&\,6\alpha_2\lambda ^4 \left[\alpha_1  Q_3  X^A  Y^A +2 (\alpha_1 +\alpha_2 ) (X^A +Y^A )+2 (\alpha_1 -\alpha_2 ) (X^A -Y^A ) \cos (2 \phi )     \right]\nonumber\\
&-12 \beta_3  \lambda ^3 Y^A  \left(4 \alpha_2 +\alpha_1  Q_3  X^A \right)\cos (\phi ) -4\lambda ^2  \beta_3^2 Q_3 Y^A Y^C \left(2+Q_1 X^A[1-\cos(2\phi)]\right),\\
M_{-2}=&\,4 \lambda ^5  \left[8 \alpha_2 +Q_3  (\alpha_1 +\alpha_2 ) \left(X^A +Y^A \right)+Q_3  (\alpha_1 -\alpha_2 ) \left(X^A -Y^A \right) \cos (2 \phi )\right]-16\lambda ^4\beta_3 Q_3  Y^A  \cos (\phi ),\\
M_{-3}=&\,32\lambda^6 Q_3 /3,
\end{align}
and finally
\begin{align}
N_0 =&\,\frac{9\lambda }{Y^C} \alpha_2  m_2  \cos (\phi )-6 \beta_3 Q_1 (m_1+m_2\cos(2\phi)),\\
N_{-1}=&\,\frac{3m_2 \lambda ^2 Q_3   \cos (\phi )}{Y^C }+\frac{2 \beta_3}{\alpha_1\alpha_2}  \lambda  Q_3  Q_1\left((\beta_1+\beta_2)+(\beta_1-\beta_2) \cos (2 \phi )\right)+\frac{12 (\beta_2-\beta_1) \lambda ^2 \cos (\phi )}{\alpha_1 X^A  Y^C },\\
N_{-2}=&\,\frac{4 (\beta_2-\beta_1) \lambda ^3 Q_3   \cos (\phi )}{\alpha_1\alpha_2X^A  Y^C }.
\end{align}

\section*{Acknowledgements}
We gratefully acknowledge helpful conversations with Michael Graham and Jean-Luc Thiffeault, and the support of the NSF/NIH (DMS-1661900) and the UW-Madison Welton Summer Sophomore Apprenticeship program.
\bibliographystyle{unsrt}
\bibliography{Bigbib}

\newpage

\begin{table}[htp]
\caption{List of mathematical symbols}
\begin{center}
\begin{tabular}{|c|c|}
\hline
Body geometry &  \\
\hline
$a$ & Semi-major axis length \\
$e$ & Eccentricity \\
\hline
Flagellum geometry &  \\
\hline
$b$, $k$, $\alpha$ & Helical amplitude, wavenumber, $\alpha=\sqrt{1-(kb)^2}$\\
$s$, $L$, $\lambda$ & Arc-length, total length, $\lambda=L/a$\\
$\e$, $\delta$ & Aspect ratio, phase\\
\hline
Helical trajectory geometry &  \\
\hline
$\kappa$, $\tau$ & Curvature, torsion \\
$A$, $\psi $ & Amplitude, pitch angle \\
\hline
Hook geometry and mechanics &  \\
\hline
$\phi$, $\phi_0$ & Bending angle, preferred angle \\
$B$, $\ell_h$ & Bending stiffness, length \\
$M_a$, $\tilde{M}_a$  & Active moment from body to hook, $\tilde{M}_a = (B/\ell_h)^{-1}M_a$\\
\hline
Reference frame &  \\
\hline
$\b{P}$ & Flagellar orientation \\
$\Uo$, $\Oo$ & Body translational, rotational velocities \\
$\Uo^f$, $\Oo^f$  & Flagellum translational, rotational velocities\\
$\b{X}$ & Hook / flagellar base position $\b{X}=a \b{e}_3$ \\
$\b{X}^f$, $\b{\hat{s}}$ & Position on flagellum, tangent vector $\b{\hat{s}}=\partial_s \b{X}^f$ \\
\hline
Lab frame &  \\
\hline
$\b{r}$ & Body centroid position\\
$\b{p}$ & Flagellar orientation, $\b{p}=\D\cdot \b{P}$ \\
$\U$, $\O$ & Body translational, rotational velocities: $\U = \dot{\b{r}}  = \D \cdot \Uo$, $\O = \D\cdot \Oo$ \\
$\O_D$ & Orthonormal basis rotational velocity, $\O_D = \O +\dot{\theta}\b{d}_3$\\
$\O^p$ & Flagellum rotational velocity, $\O^p=\O_D+\dot{\phi} \b{d}_2$ \\
$\b{x}$ & Flagellar base position $\b{x}=\b{r}+a \b{d}_3$ \\
$\dot{\theta}$, $\omega$ & Flagellar precession rate, axial flagellum spin rate\\
\hline
Body hydrodynamics &  \\
\hline
$X^A$, $Y^A$, $X^C$, $Y^C$& Resistance coefficients \\
$\b{F}_0(\b{P})$, $\b{M}_0(\b{P})$ & Total viscous force, torque (reference frame) \\
\hline
Flagellum hydrodynamics &  \\
\hline
$\b{f}_0(s)$, $\b{u}_0(s)$ & Viscous force/length, local velocity (reference frame) \\
$\b{F}^f_0$, $\b{M}^f_0$ & Total viscous force, torque (reference frame) \\
$c$ & $c=\log(1/\e^2)-1$ \\
$\{\Atens,\, \Btens,\, \Dtens\}$, $\{\tilde{\Atens}, \, \tilde{\Btens},\, \tilde{\Dtens}\}$ & Resistance tensors, mobility tensors\\
$\{\xi_i,\, \eta_i,\, \zeta_i\}$, $\{\alpha_i, \, \beta_i,\, \gamma_i\}$ & Resistance coefficients, mobility coefficients\\
\hline
\end{tabular}
\end{center}
\label{default}
\end{table}%
 \end{document}